\documentclass[a4paper,fleqn,usenatbib]{mnras}

\usepackage[T1]{fontenc}
\usepackage{ae,aecompl}
\usepackage{rotating}

\usepackage{graphicx}	
\usepackage{amsmath}	
\usepackage{amssymb}	

\title[Flare stars in Upper Sco]{The Flare-Activity of
  2MASS\,J16111534-1757214 in the Upper Scorpius association
  \footnote{Partly based on observations made with ESO Telescopes 
  at the La Silla Paranal Observatory under programme  097.C-0040(A).}}

\author[Eike W. Guenther et al.]
  {E.W. Guenther, $^{1}$\thanks{E-mail: guenther@tls-tautenburg.de}
  D. W\"ockel, $^{1}$,
  P. Chaturvedi, $^{1}$,
  V. Kumar, $^{2,3}$,
  M.K. Srivastava, $^{2}$, 
    \newauthor
  P. Muheki $^{1,4}$ \\
$^{1}$ Th\"uringer Landessternwarte Tautenburg, Sternwarte 5, 07778
  Tautenburg, Germany \\
$^{2}$ Astronomy \& Astrophysics Division, Physical Research Laboratory, Ahmedabad
380009, India \\
$^{3}$ Indian Institute of Technology, Gandhinagar 382335, India \\
$^{4}$ Mbarara University of Science and Technology, Department of
  Physics, Uganda \\ 
}

\date{Accepted XXX. Received xxxx 2021; in original form 4 December 2020}

\pubyear{2020}

\begin{document}
\label{firstpage}
\pagerange{\pageref{firstpage}--\pageref{lastpage}}
\maketitle


\begin{abstract}
Flares are known to play an important role for the evolution of the
atmospheres of young planets. In order to understand the evolution of
planets, it is thus important to study the flare-activity of young
stars.  This is particularly the case for young M-stars, because they
are very active.  We study photometrically and spectroscopically the
highly active M-star 2MASS\,J16111534-1757214. We show that it is a
member of the Upper Sco OB association, which has an age of 5-10
Myrs. We also re-evaluate the status of other bona-fide M-stars in
this region and identify 42 members. Analyzing the K2-light curves, we
find that 2MASS\,J16111534-1757214 has, on average, one super-flare with
$\rm E\geq 10^{35}$\,erg every 620 hours, and one with $\rm E\geq
10^{34}$\,erg every 52 hours. Although this is the most active M-star
in the Upper Sco association, the power-law index of its
flare-distribution is similar to that of other M-stars in this
region. 2MASS\,J16111534-1757214 as well as other M-stars in this
  region show a broken power-law distribution in the flare-frequency
  diagram.  Flares larger than $\rm E \geq 3\,10^{34}$\,erg have a
  power-law index $\rm \beta=-1.3\pm0.1$ and flares smaller than that
  $\rm \beta=-0.8\pm0.1$. We furthermore conclude that the flare-energy
  distribution for young M-stars is not that different from
  solar-like stars.
\end{abstract}
  




\section{Introduction}
\label{sectI}

Studies of planets in the mass range between 1 and 15 $\rm M_{Earth}$
have revealed a large diversity of their densities.  Planets in this
mass-range can have densities as low as $\rm 0.05\,g\,cm^{-3}$ like
Kepler-51 \citep{masuda14}, or as high as $\rm 13\,g\,cm^{-3}$, like
K2-106 \citep{guenther17} . This large diversity in densities is
related to different compositions of the planets. High-density planets
must be bare rocks and the low-density ones must have extended
Hydrogen-dominated atmospheres. Recent studies have revealed that
close-in planets with radii smaller than 1.4 $\rm R_{Earth}$ are
rocky, and planets larger than 1.8 $\rm R_{Earth}$ have extended,
Hydrogen dominated atmospheres
\citep{owen2013,jin2014,lopez2014,fridlund2020}.  Only few planets
have radii between 1.4 and 1.8 $\rm R_{Earth}$.
 
There are three possible mechanisms that can explain why some planets
have a Hydrogen dominated atmosphere and others not: 1.) gas-poor
formation scenario \citep{owen2013}; 2.) atmospheric losses driven by
the energy release from the formation-process \citep{ginzburg2018,
  gupta2019, gupta2020}, and 3.) atmospheric losses due to the
XUV-radiation from the host star \citep{lammer2014, linsky2015a}.  The
XUV-radiation is the soft X-ray plus the extreme
UV-radiation\footnote{NUV: 1700 to 3200 \AA \,(3.875 - 7.293 eV), FUV:
  912 to 1700 \AA \,(7.293 - 13.59 eV), EUV: 100 to 912 \AA \,
  (13.59-124 eV), soft X-ray: 2-100 \AA \,(6.4 keV - 124 eV), XUV:
  2-912 \AA \, (13.59 eV - 6.4 keV).}.

Although there are currently some indications that gas-poor formation
might be the dominant process for planets of late K- and M-stars
\citep{cloutier2020}, atmospheric erosion exists and must be taken in
to account.  All young, low-mass stars emit XUV-radiation, thus all
young planets with atmospheres must have some XUV-driven mass-loss.
We thus have to study this process in any case. XUV-radiation from the
host star ionizes and heats up the outer layers of the planet which
then escapes \citep{linsky2015a}.  Particularly important are studies
of the atmospheric escape-rates in M-stars, because M-stars are the
preferred targets for the search of low-mass planets due to their
abundance in the galaxy, and because it is comparatively easy to
detect low-mass planets orbiting them \citep{quirrenbach2020}.
However, potentially habitable planets orbit close to the host stars
and the high activity phase lasts longer for M-stars than for
solar-like stars, so that they are exposed to a high level of
XUV-radiation over a long time \citep{johnstone2015,johnstone2020}.

Flares contribute significantly to the XUV-radiation, and the younger
and more active the star is, the larger the contribution from flares
to the XUV-radiation \citep{linsky2015a, telleschi2005} .  The coronal
emission measure distribution can be reproduced assuming that it is
the result of a superposition of stochastically occurring flares
\citep{telleschi2005, wood2018}. Flares are hotter than stellar
coronae.  The XUV-radiation from flares thus penetrates deeper into
the atmospheres of the planets.

Because of their sporadic nature, long monitoring campaigns are needed
to find out how many flares with what energy are emitted. The key
question is: Is the contribution from a few super-flares larger, or
smaller than the contribution of the many small flares? In other
words: Is $\rm\beta$ smaller or larger than minus one (see
Section~\ref{sectVb})? If small flares dominate, a typical measurement
of the XUV-flux will be the sum of the emission from the corona plus
all the small flares that appeared during the exposure. If rare but
large flares dominate, one measurement of the XUV-flux is not enough.

In our previous works, we have studied the two highly active M-stars
AD\,Leo and EV\,Lac \citep{muheki2020a, muheki2020b}.  Since AD\,Leo
and EV\, Lac have ages of about 250-300 Myr \citep{shkolnik2009}, the
next logical step is to study the flare activity of M-stars that are
significantly younger.  Recently \citet{ilin2020} studied the
light-curves of 2111 members of open clusters have ages between
135 Myrs and 3.6 Gyrs.  They find a rapid decline of the flare
activity for M1-M2 stars when the star have spun down the rotation
rate to 10d at the age of about 700 Myrs.

M-stars in the Upper Scorpius OB association are ideal targets for
such a study, because Upper Sco has an age of 5-10 Myrs
\citep{fang2017, david2019}.  As we will show in this article 90\% of
the M-stars in Upper Sco have rotation periods of less than 10d and
are thus in the high activity phase. Furthermore, planets have also
been discovered in this region.  K2-33 (EPIC\,205117205) is an M-star
that has a transiting planet of $\rm R_p=5.04_{-0.37}^{+0.34}
R_\oplus$ with an orbital period of 5.425 days \citep{mann2016}.
Depending on the mass of the planet and the activity level of the
star, the atmosphere could be subject to extremely high escape rates
\citep{kubyshkina2018}. Perhaps, we witness the transition of a
mini-Neptune to a rocky planet.  Very recently another star hosting
two planets has been found \citep{bohn2020}.  It is thus certain that
planets have already formed in this region. The region has also been
observed in the Kepler K2 mission, which provides us with a good
data-set to study flares.

Unfortunately, the K2 observations of K2-33 do not allow to determine
the power-law of the flare-frequency distribution accurately enough to
calculate atmospheric erosion of the planet due to the XUV-radiation
from flares.  We thus take a slightly different approach: We identify
the most active M-star in Upper Sco, determine the power-law index of
the flare distribution $\rm \beta$. In the next step we determine the
average $\rm \beta$ for all M-stars in Upper Sco, compare it to this
star.  If the two are the same, we know that this relation is
universal for M-stars at that age and can be used for modeling the
erosion rate of M-star planets at that age.  The critical question is:
Is $\rm\beta$ smaller, or larger than minus one for active and less
active M-stars at this ages.

The most active M-star in Upper Sco is 2MASS\,J16111534-1757214
(EPIC\,205375290; called 2M1611-1757 from now on).  This star is also
interesting because it is the first M star where solar-like
oscillations were discovered \citep{muellner2020}.  Identifying the
most active M-star in Upper Sco is also important for another reason:
What we need is the energy emitted by flares in the XUV-regime, not
the optical. For that we have to observe flares simultaneously in the
optical and in X-rays, which can only be done for a star with a very
high flare-rate. Coronal-Mass Ejections (CMEs) can also be
important for the erosion of planetary atmospheres. CMEs are rare, we
only have a chance to detect them if we have identified the most
active young M-star.

A first estimate of how much energy is released in the XUV can already
be obtained using the branching ratio for flares on other stars. This
allows us to answer the question whether flares can be important for
the erosion of planetary atmospheres, or not.

The mass and radius and other properties of 2M1611-1757 are given in
Section~\ref{sectII}.  In Section~\ref{sectIII} we show that
2M1611-1757 is a member of the Upper Sco association. The
spectroscopic and photometric flare studies and the results obtained
from them are discussed in Section~\ref{sectIV} and
Section~\ref{sectV}.  In Section~\ref{sectV} we furthermore show that
the $\rm \beta$-value of 2M1611-1757 is similar to other M-stars in
this region. The impact of flares on planetary atmospheres is
discussed in Section~\ref{sectVI}. Conclusions are in
Section~\ref{sectVII}.

\section{Mass, radius and other properties of 2M1611-1757}
\label{sectII}

Thanks to the Gaia mission, it is now possible to determine accurate
stellar radii by combining the parallax measurements with the
measurements of the relative brightness. Because of the extinction,
photometric measurements at infrared wavelengths are preferred.

\begin{figure}
\includegraphics[height=0.35\textheight,angle=-90.0]{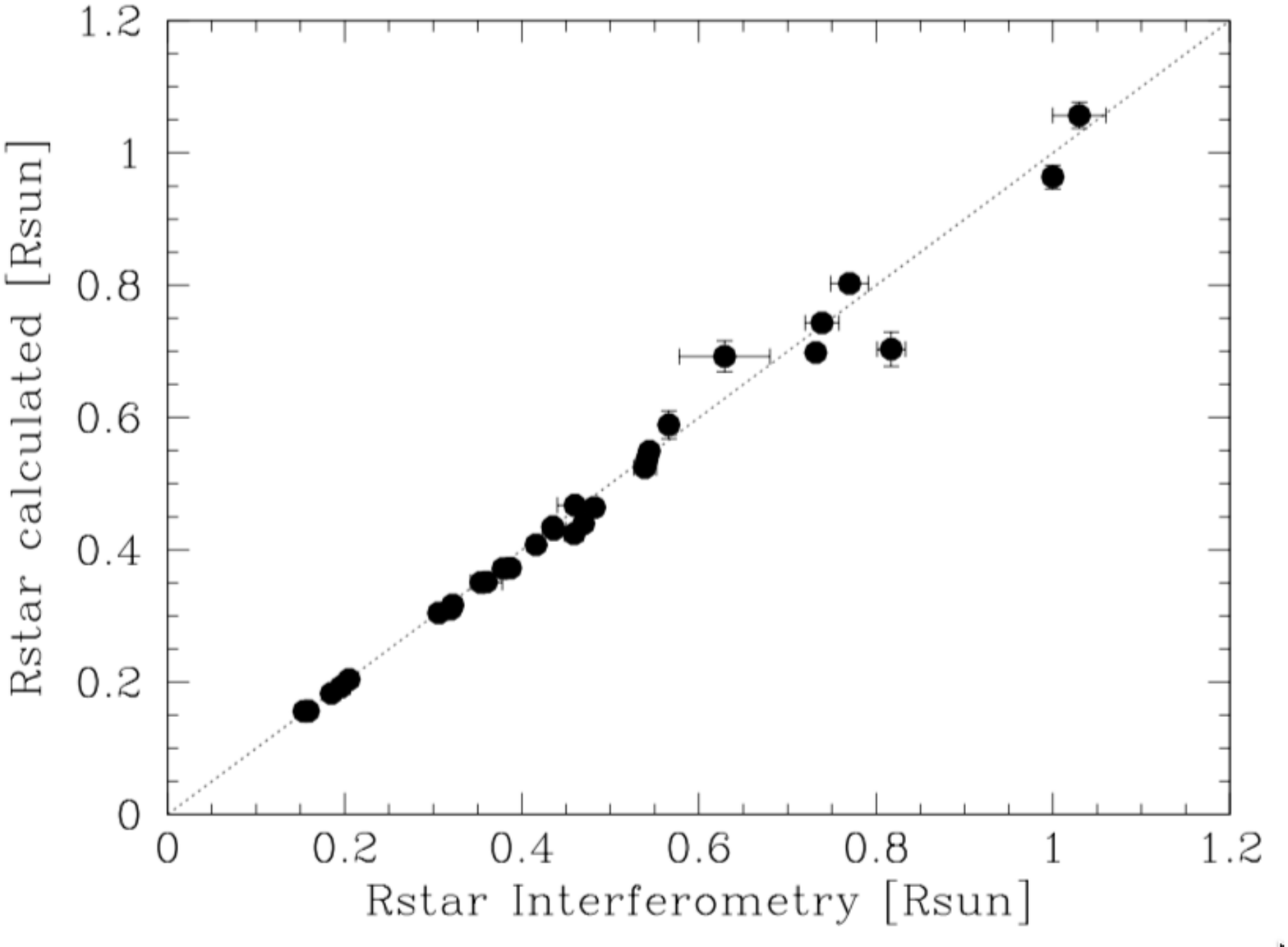}
\caption{Comparing measured and calculated diameters of stars using
  the method described in the text. The error is on average 
  0.021 $\rm R_*$. }
\label{Fig06}

\end{figure}

Infrared photometry has additionally the advantage that there are less
spectral lines than in the optical. Because M-stars have less
absorption lines in the infrared, the correction for the
line-blanketing is smaller. Figure \ref{Fig06} shows a comparison
between stellar diameters determined interferometrically and
calculated using 2\,MASS-photometry and the parallaxes from the early
release of the Gaia DR3 cathalog \citep{2MASS, gaia16, gaia18,
  gaia20}.

The error is on average 0.021 $\rm R_*$.  Using this method, we obtain
$\rm R_*=1.247\pm0.096\,R_{sun}$ for 2\,MASS\,1611-1757. Using the
evolutionary models published by \citet{baraffe2015} for an age of 5
and 10 Myrs and the absolute brightness of the star at infrared
wavelength, we derive a mass of $\rm M_*=0.71\pm0.08\,M_{sun}$ for
this star.  The error of the mass is dominated by the error of the
age. The mass and radius derived by us are thus in excellent agreement
with previous determinations. All parameters for 2M1611-1757 from the
literature and derived by us are given Table~\ref{tab01}.

\begin{table}
\caption{Parameters of the star 2M1611-1757}
  \begin{tabular}{l l }
\hline
Parameter    & Value \\
\hline
Name               & 2MASS\,J16111534-1757214, \\
                   & EPIC\,205375290, \\ 
                   & 1RXS\,J161115.1-175741, \\ 
                   & Gaia\,EDR3\,6249000566108106112 \\ 
RA[h:m:s]          &  16:11:15.3443 $^{(3)}$ \\ 
DE [d:m:s]         & -17:57:21.42779  $^{(3)}$ \\ 
RA[d]              &  242.8139346 \\
DE[d]              & -17.95595216 \\
Glon [deg]         & $355.6973728700222^{(3)}$ \\
Glat [deg]         & $+23.83794726211355^{(3)}$ \\
pm-RA[mas/yr]      & $-8.954\pm0.022^{(3)}$   \\ 
pm-DE[mas/yr]      & $-24.626\pm0.016^{(3)}$   \\ 
RV [km/s]          & $-7.23\pm0.17^{(1)}$   \\
plx [mas]          & $7.364\pm0.017^{(3)}$   \\ 
Distance [pc]      & $135.78\pm0.32^{(3)}$\\
Distance module [mag] & $5.6642\pm0.0051$  \\ 
Mass [M$_\odot$]   & $0.6\pm0.1^8$, $0.71\pm0.08^{(5)}$   \\
Radius [R$_\odot$] & $1.24^{(3)}$, $1.247\pm0.096^{(5)}$ \\
SpecType           & $\rm M1^{(4,5)}$ \\
Teff[K]            & $3535^{(3)}$, $3750^{(4)}$ , $3670\pm180^{(12)}$ \\
Lum. [L$_\odot$]   & $0.215^{(3)}$, $0.38^{(4)}$, $0.29\pm0.03^{(8)}$ \\
Age [Myr]          & 1.3-5.5$^{(8)}$, 5-10$^{(9)}$ \\
Rot. period [d]    & $6.0308\pm0.0084^{(5)}$  \\
BP [mag]           & $14.1204\pm0.080^{(3)}$ \\ 
RP [mag]           & $11.8219\pm0.056^{(3)}$ \\ 
V [mag]            & 13.3$^{(1)}$ \\
R [mag]            & 11.9$^{(1)}$ \\
G [mag]            & $12.9300\pm0.0023^{(1)}$ \\
Kp [mag]           & 13.233  \\
J [mag]            & $10.227\pm0.027^{(2)}$ \\
H [mag]            & $9.486\pm0.023^{(2)}$ \\
K [mag]            & $9.204\pm0.019^{(2)}$ \\
W1 (3-4 $\mu m$) [mag]   & $8.754\pm0.023^{(7)}$ \\
W2 (4-8 $\mu m$) [mag]   & $8.250\pm0.020^{(7)}$ \\
W3 (8-15 $\mu m$) [mag]  & $6.291\pm0.016^{(7)}$ \\
W4 (15-30 $\mu m$) [mag] & $4.723\pm0.029^{(7)}$ \\
$\rm Fmm\lambda $ [$\rm mJy^{mm}$] & $<0.18^{0.88}$\,$^{(8)}$ \\
Av [mag]           & $1.6^{(4)}$ \\
$\rm EW\,LiI$ [\AA]       & $0.600\pm0.035^{(5)}$ \\
$\rm EW\,H{\alpha}$ [\AA] & $-2.4^{(4)}$, $-4.7\pm0.1^{(5)}$ \\
$\rm EW\,H{\beta}$ [\AA]  & $-1.4\pm0.2^{(5)}$ \\
$\rm F_{H{\alpha}}$ [$\rm 10^{29}\,erg\,s^{-1}$] & $3.29\pm0.04^{(10)}$, \\
                                           & $2.94\pm0.28^{11}$, $2.58\pm0.16^{(11)}$ \\
$\rm F_{H\,{\beta}}$ [$\rm 10^{29}\,erg\,s^{-1}$] & $0.47\pm0.01^{(10)}$ \\
$\rm L_x $  [$\rm 10^{29}\, erg\,s^{-1}$]         & 16 \\
HR1                      & $0.96\pm0.39^{(6)}$ \\
HR2                      & $0.44\pm0.29^{(6)}$ \\
X-ray [ct/s]             & $0.0467\pm0.0135^{(6)}$ \\
\hline
\end{tabular}
\label{tab01}
\\
$^1$ SIMBAD, Centre de donn\'e es astronomiques de Strasbourg \citep{wenger00}.\\
$^2$ \cite{2MASS}.\\
$^3$ Early release of Gaia DR3 (\url{https://www.cosmos.esa.int/gaia})\citep{gaia16, gaia18, gaia20}.\\
$^4$ \cite{preibisch01}.\\
$^5$ This work \\
$^6$ \cite{ROSAT2000}.\\
$^7$ AllWISE Data Release \citep{cutri2013}.\\
$^8$ \cite{garufi20}. \\
$^9$ \cite{fang2017}. \\
$^{10}$ This work, FLAMES spectra \\
$^{11}$ This work, Mt. Abu spectra \\
$^{12}$ \cite{muellner2020} \\
\end{table}

\section{2M1611-1757 is a member of the Upper Scorpius OB association}
\label{sectIII}

Using the 2dF multi-object spectrograph at the Anglo-Australian
Telescope (AAT), \cite{preibisch01} observed 6 $\rm deg^2$ area in the
Upper Sco OB association and identified 98 bona-fide members of it.
The AAT spectra had a resolution of 1.8\,\AA\, ( $\rm \Delta \lambda /
\lambda =4000$) and cover the wavelength range from 6150 to 7250\,\AA .

\begin{figure}
\includegraphics[height=0.25\textheight,angle=0.0]{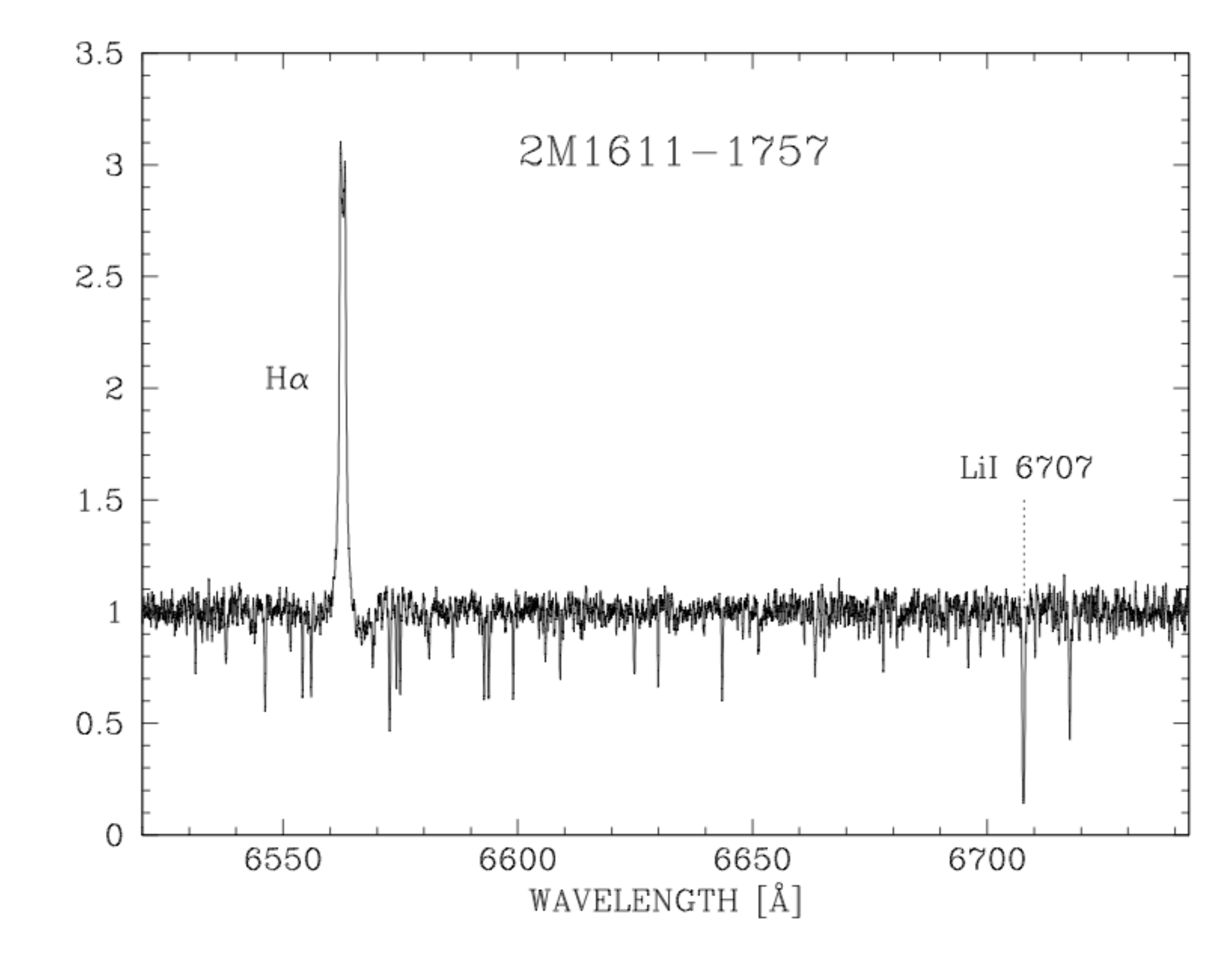}
\caption{Part of the normalized FLAMES spectrum of 2M1611-1757 showing the
$\rm H\alpha$ and the $\rm\ion{Li}{i}\,6707$\AA\, line. }
\label{Fig04}
\end{figure}

Because of the dense forest of spectral-lines in M-stars, it was not
always easy to measure the equivalent widths of the
$\rm\ion{Li}{i}\,6707$ lines in the AAT-spectra.  Using the
FLAMES-UVES spectrograph at the ESO VLT UT2 telescope in ESO programe
097.C-0040(A) we thus re-observed 57 stars in this region.  The FLAMES
spectra have a resolution of $\rm \Delta \lambda / \lambda =47000$ and
cover the wavelength range from 4820 to 5790 \AA \, and 5880 to
6840\,\AA . The higher resolution of the FLAMES spectra made it much
easier to measure the equivalent width of the
$\rm\ion{Li}{i}\,6707$\,\AA\,line. Fig. \ref{Fig04} shows part of the
spectrum containing the $\rm H\alpha$ and the
$\rm\ion{Li}{i}\,6707$\,\AA\, lines. The spectrum was taken on 1 June
2016 from UT 04:06 to UT 04:36 (HJD 2457540.67688 to 2457540.69748).

Combining these measurements with the parallax measurements from Gaia,
we found that 42 of the stars are members of Upper Sco association but
16 are not.  The results are presented in Tab.~2 and Tab.~3. The
average parallax of the confirmed members in Upper Sco is
$6.87\pm0.36$ mas, ($\rm d=145.6\pm7.6\,pc$). The average proper
motions of confirmed members are $\rm pm-RA=-10.6\pm6.0$\,mas and $\rm
pm-DEC=-23.0\pm1.4$\,mas, respectively. The values obtained for
2M1611-1757 are thus consistent with a membership in Upper Sco.

\begin{figure}
\includegraphics[height=0.29\textheight,angle=0.0]{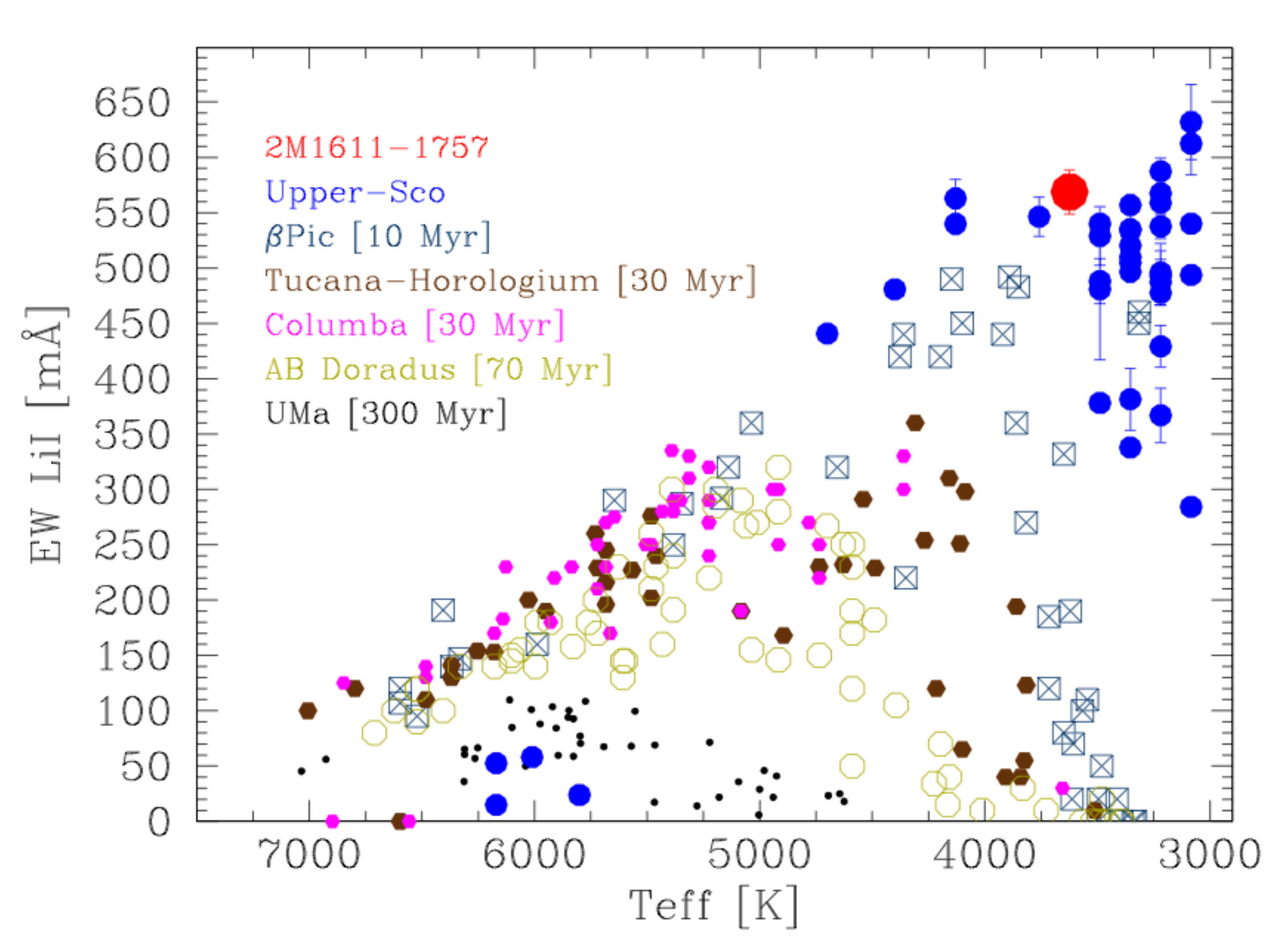}
\caption{Equivalent width of the LiI\,6707\AA\, lines in clusters of 
different age. The big red dot is 2M1611-1757.}
\label{Fig11}
\end{figure}

We also determined the rotation-rate of the stars from the
light-curves. We find that 31 of the 36 M-stars with known rotation
rates have periods shorter than 10d.  That means 86\% (74\%) of the
M-stars in this region are in the high activity phase. We did not
observe K2-33 with FLAMES, because the results for this star were
already published by \citet{mann2016}. With a rotation period of $\rm
6.29\pm0.17$\,days it belongs to the class of active M-stars.  Figure
\ref{Fig11} shows the equivalent width of
$\rm\ion{Li}{i}\,6707$\,\AA\, line of the stars in Upper Sco together
with the equivalent width of other young clusters. The equivalent
width of the $\rm\ion{Li}{i}\,6707$\,\AA-line for 2M1611-1757 is $\rm
EW=600\pm35$ m\AA\, demonstrating that it is a member of the
cluster. In some cases the S/N of the spectra was too low to detect
the $\rm\ion{Li}{i}\,6707$\,\AA-line.  The stars that neither have a
significant $\rm\ion{Li}{i}\,6707$\,\AA-line, nor the right distance
are considered not to be members of Upper Sco.

The X-ray brightness of 2M1611-1757 is $\rm log(L_x)=30.2$ $\rm
log(erg\,s^{-1})$. For comparison: The Sun has $\rm log(L_x)=$
26.4-27.7 $\rm log(erg\,s^{-1})$ and solar-like stars in the Pleiades
$\rm log(L_x)=$ 29.1-29.6 $\rm log(erg\,s^{-1})$ \citep{giardino2008}.
The X-ray flux thus is 300-6000 times larger than that of the Sun.
The large X-ray brightness, the distance, the equivalent width of the
LiI\,6707 line, and the extinction all support the hypothesis that
2M1611-1757 is an M-star in the Upper Sco OB association.
AO-observations with VLT/SPHERE show that 2M1611-1757 is not a visual
binary. No disk was detected with ALMA \citep{garufi20}.

\begin{table*}
\caption{Stars that are not members of Upper Sco}
  \begin{tabular}{l l r c c c r r r}
\hline
EPIC & ST &  RA    &  DEC  & $\rm Kp^1$ & $\rm G^2$ & $\rm plx^3$ & $\rm p.m.^3$ & $\rm p.m.^3$ \\
     &    & h:m:s  & d:m:s & [mag]      & [mag]     & [mas]       & RA           & DEC     \\
     &    &        &       &            &           &             & [mas/a]      & [mas/a] \\
\hline
204809256 & K3 & 16:07:40.908 & -20:45:54.23 & 14.84 & 14.83 & $0.190\pm0.040$ & $-9.751\pm0.039$  & $-8.236\pm0.027$ \\
204820565 & K1 & 16:07:47.312 & -20:42:52.39 & 13.98 & 15.68 & $0.964\pm0.051$ & $6.137\pm0.059$   & $1.508\pm0.038$ \\
204826968$^4$ & -  & 16:07:18.264 & -20:41:08.08 & 13.16 & 13.46 & $1.127\pm0.064$ & $-1.149\pm0.070$  & $-52.730\pm0.047$ \\
204873961 & K4 & 16:08:51.400 & -20:28:25.02 & 13.91 & 13.92 & $0.395\pm0.020$  & $-4.942\pm0.027$  & $-3.169\pm0.017$ \\
204967795 & K8 & 16:01:51.816 & -20:02:19.31 & 12.53 & 12.49 & $11.423\pm0.017$ & $75.354\pm0.019$  & $-21.651\pm0.011$ \\
205027701 & K3 & 16:11:02.864 & -19:45:18.36 & 13.10 & 13.11 & $0.085\pm0.014$  & $-2.460\pm0.017$  & $-5.300\pm0.012$ \\
205040048 & G8 & 16:08:40.438 & -19:41:45.04 & 14.99 & 19.19 & $0.432\pm0.311$  & $3.871\pm0.377$   & $-6.424\pm0.284$ \\
205065331 & K6 & 16:07:25.968 & -19:34:27.57 & 14.52 & 14.45 & $0.136\pm0.030$  & $-4.432\pm0.032$  & $-7.845\pm0.026$ \\
205068387 & K4 & 16:11:00.071 & -19:33:36.41 & 13.26 & 13.24 & $0.341\pm0.016$  & $-5.635\pm0.022$  & $-4.211\pm0.013$ \\
205082091 & K5 & 16:07:50.956 & -19:29:34.75 & 13.24 & 13.19 & $0.226\pm0.018$  & $-1.501\pm0.020$  & $-4.810\pm0.017$ \\
205084272 & F8 & 16:07:51.787 & -19:28:55.18 & 12.80 & 12.80 & $4.340\pm0.043$  & $-28.000\pm0.044$ & $-18.676\pm0.033$ \\
205085384 & K1 & 16:09:39.751 & -19:28:35.55 & 13.01 & 12.99 & $1.632\pm0.035$  & $-16.543\pm0.039$ & $-15.473\pm0.031$ \\
205137523 & F8 & 16:10:23.563 & -19:13:04.37 & 12.50 & 12.56 & $2.283\pm0.026$  & $-3.964\pm0.034$  & $-30.466\pm0.024$ \\
205158932 & K4 & 16:10:57.684 & -19:06:35.24 & 13.80 & 13.72 & $1.147\pm0.019$  & $2.104\pm0.022$   & $-5.020\pm0.016$ \\
205160903 & K4 & 16:08:31.682 & -19:05:58.83 & 13.13 & 13.14 & $0.359\pm0.019$  & $0.240\pm0.019$   & $1.817\pm0.013$ \\
205168450 & K6 & 16:09:09.907 & -19:03:41.05 & 11.94 & 11.91 & $0.498\pm0.015$  & $-3.000\pm0.019$  & $-4.337\pm0.013$ \\
\hline
\end{tabular}
\label{tab03}
\\
$^1$ Kepler magnitude, $^2$ Gaia magnitude, $^3$ Gaia (\url{https://www.cosmos.esa.int/gaia})\citep{gaia16, gaia18, gaia20}, \\
$^4$ two stars within 3 arcsec\\
\end{table*}

\section{Spectroscopic monitoring}
\label{sectIV}

\begin{figure}
\includegraphics[height=0.25\textheight,angle=0.0]{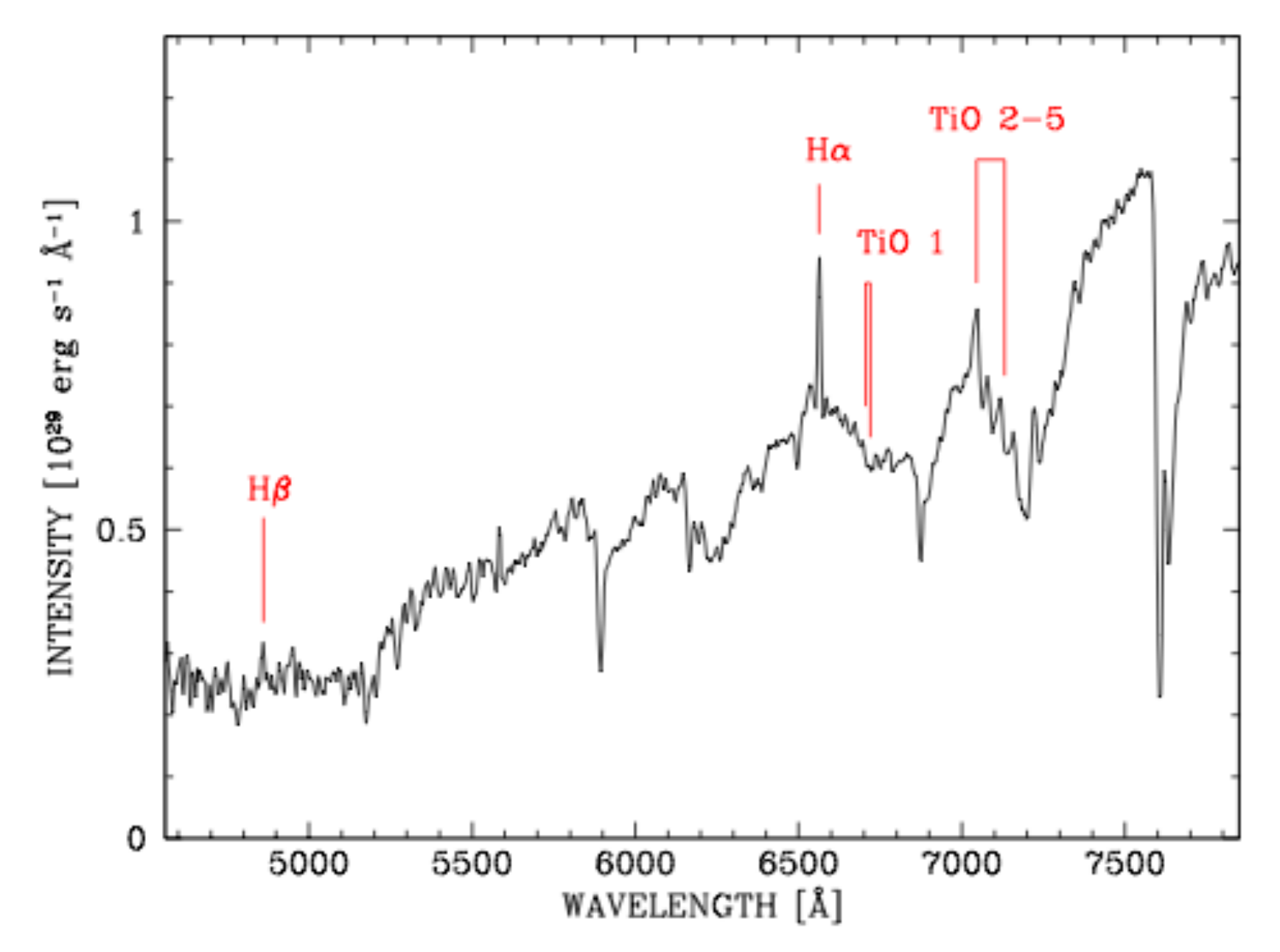}
\caption{Low-resolution spectrum of 2M1611-1757. The intensity-scale 
is the total emission from the star. Prominent features in the spectrum
of the star are marked. }
\label{Fig05}
\end{figure}

We observed 2M1611-1757 in two observing nights using the Mt. Abu
Faint Object Spectrograph and Camera - Pathfinder (MFOSC-P) mounted on
the Physical Research Laboratory (PRL) 1.2m telescope at Mt. Abu India
\citep{srivastava2018, srivastava2021}.  For our observations we used
the grating with 150 lines per mm which gives a resolution of $\rm
\Delta \lambda / \lambda \sim 500$. The spectra were exposed for
600\,s. They cover the wavelength region from 4563 to 8409 \AA.  The
observations of the first night were taken on the $\rm 5^{th}$ of May
2020 from 18:34 to 23:30 UT (HJD 2458971.27898 to 2458971.48454).  In
this night, we obtained 21 spectra.  The observations of the second
night were taken on the $\rm 20^{th}$ of May 2020 from 16:49 to 22:23
UT (HJD 2458990.20651 - 2458990.43846).  In the second night we
obtained 28 spectra. We thus monitored the star spectroscopically for
10.5 hours in total.  The detailed description of the MFOSC-P
instrument and the data reduction are given in
\cite{rajpurohit2020}. Figure\,\ref{Fig05} shows an average spectrum
of the star.  The intensity-scale is the total emission from the star.
The $\rm H\alpha$, $\rm H\beta$ and the TiO-band head are marked.  The
nomenclature is the same as in \citep{reid1995}.

\begin{figure}
\includegraphics[height=0.25\textheight,angle=0.0]{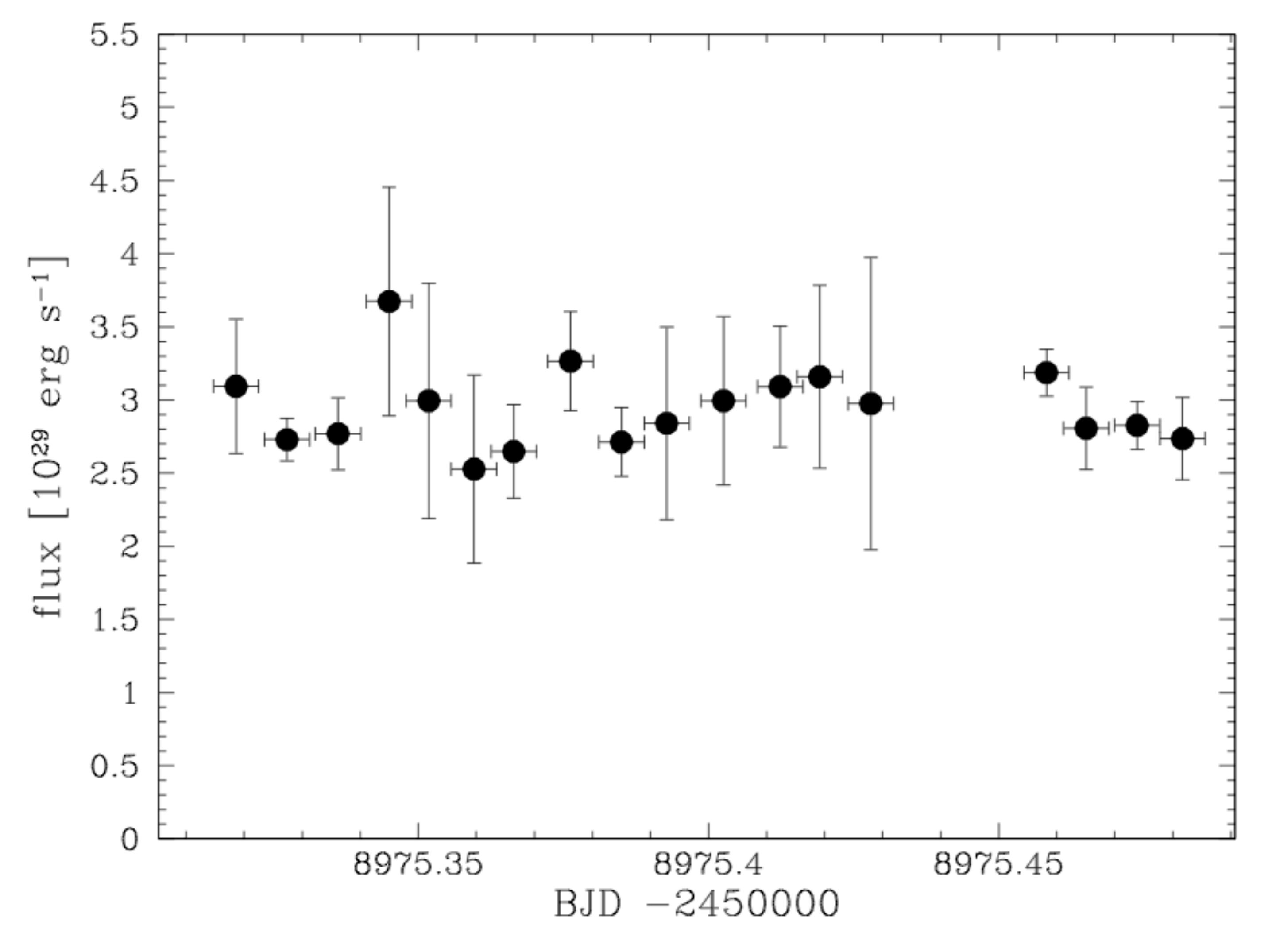}
\caption{Flux of $\rm H\alpha$ in the first night.}
\label{Fig07}
\end{figure}

\begin{figure}
\includegraphics[height=0.25\textheight,angle=0.0]{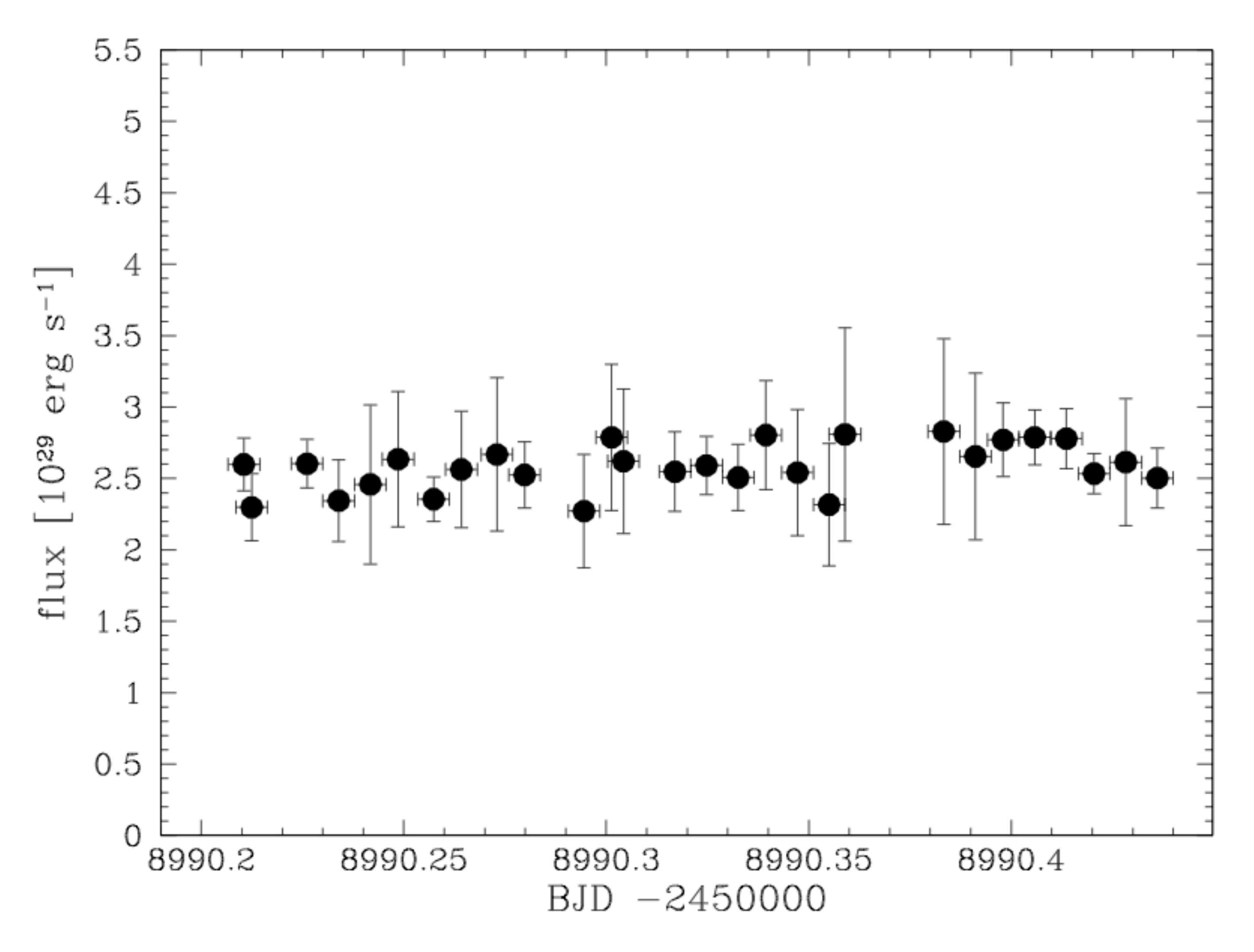}
\caption{Flux of $\rm H\alpha$ in the second night.}
\label{Fig08}
\end{figure}

The fluxes in $\rm H\alpha$ in the first and second nights are showns
in Fig.~\ref{Fig07} and Fig.~\ref{Fig08}.  The average $\rm H\alpha$
line-flux in the first night was $\rm
(2.94\pm0.06)\,10^{29}\,erg\,s^{-1}$ and $\rm
(2.58\pm0.03)\,10^{29}\,erg\,s^{-1}$ in the second.  The upper limits
of the $\rm H\alpha$-emission from flares are $\rm 4.4\,10^{31}\,erg$
in the first night and $\rm 1.4\,10^{31}\,erg$ in the second.

Using the branching ratio between the continuum in the TESS-band and
$\rm H\alpha$ emission of $\rm F_{TESS}/F_{H\alpha}=10.408\pm0.026$
from \citep{muheki2020b} we can convert these upper limits into
continuum fluxes.  We have to take into account that the wavelength
regions of TESS and Kepler are different, though. Using the typical
temperatures of the optical continuum emission of 10000 to 20000 K for
flares, this correction factor is $\rm
F_{Kepler}/F_{TESS}=1.8\pm0.2$. The upper limits $\rm
H\alpha$-emission correspond to $\rm 7-9\,10^{32}$\,erg and $\rm
2-3\,10^{32}$\,erg in the optical continuum, respectively.  The
detection limits of the spectroscopic observations are thus comparable
to the photometric observations of the Kepler-K2 mission discussed in
Section\,\ref{sectV}.

Large spots on the stellar surface lead to a decrease of the average
temperature of the star. Using the TiO-5 index from \citet{reid1995}
we derive a temperature difference of the star in the two nights of
$\rm \Delta T_{eff} = 90\pm69$\,K, which is insignificant. There is
thus no evidence that the spot-coverage was much larger in one night
than in the other.

\section{Photometric monitoring: The Kepler-K2 light-curves}
\label{sectV}

Based on the list of stars in the Upper Sco association from
\citet{preibisch01} we put in a proposal to observe 119 stars in this
region in the Kepler K2-mission. The Kepler satellite observed these
stars continuously for 1860 hours. In total 3231 photometric
measurements were obtained.  The time sampling is one photometric
measurement every 34.5 minutes.

\subsection{Rotation rate and filling factor}
\label{sectVa}

\begin{figure}
\includegraphics[height=0.25\textheight,angle=0.0]{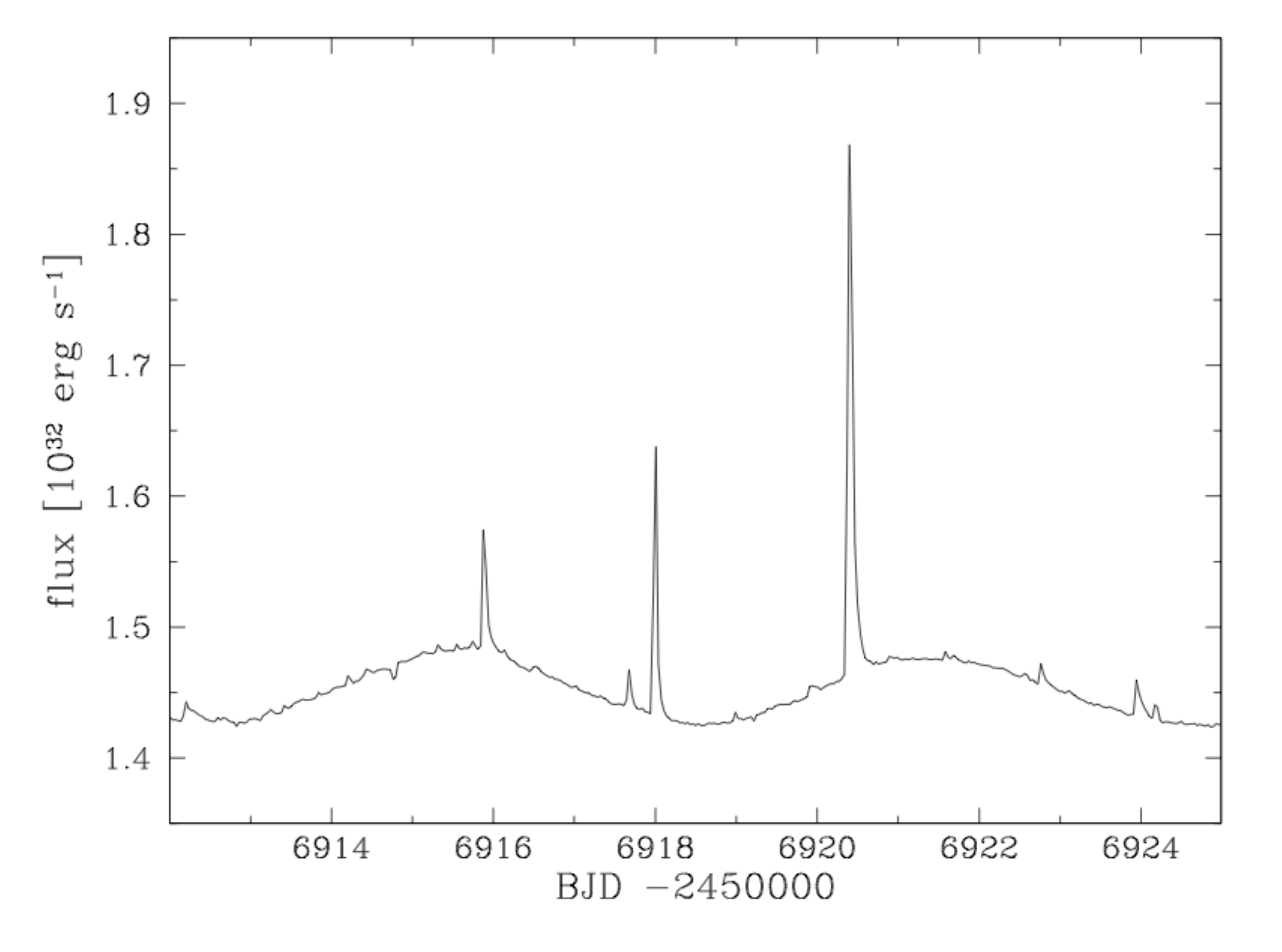}
\caption{An extract of the light-curve showing numerous flares and the
  modulation due to star-spots }
\label{Fig01}
\end{figure}

Stellar spots that are not located at the poles cause as a sinusoidal
modulation of the light-curve (Fig.~\ref{Fig01}), which allows to
determine the rotation period of a star. From the period of this
modulation we obtain a rotation period of $\rm 6.0308\pm0.0084$ days.

The average depth of minima in the light-curve is $\rm 3.45\pm0.29\%$.
The depth of the minima allows to calculate the filling-factor of the
spots that are not located at the poles. Because many active stars
have polar spots, the filling-factors derived are lower limits.

Using the relation
$${\rm T_{spot} = T_{star} -3.58 \times 10^{-5} T_{star}^{2} -
0.249\times\,T_{star}+808\,[K]}, $$
gives a spot temperature of $\rm 3093\pm20$\,K \citep{notsu2019}.
The lower limit of spot-filling factor is $\rm 7.1\pm0.9$ \% as derived
from the relation \citep{jackson2013}:
$$ {\rm A_{spot}/A_{star} = \Delta F/F
  \,[1-(T_{spot}/T_{star})^4]^{-1}}.$$

\subsection{The frequency of flares in 2M1611-1757}
\label{sectVb}

During the 78 days of observations, we detected 105 flares on
2M1611-1757. Part of the K2 light-curve is shown in
Fig.~\ref{Fig01}. Clearly seen are several flares with their
characteristic rapid increase and exponential decrease.  The largest
flare had 4.0 $\rm 10^{35}$\,erg, and the smallest one detected $\rm
8\,10^{32}$ erg, in the Kepler band (4200-9000 \AA ).  Statistically,
there is one flare with $\rm E\geq 10^{35} erg$ every 620 hours, one
of $\rm E\geq 10^{34} erg$ every 52 hours, and one of $\rm E\geq
3\,10^{33} erg$ every 24 hours.  For comparison, the large solar flare
observed by Carrington in 1859 is estimate to have emitted $\rm \sim
10^{32} erg$ \citep{tsurutani2003}. Flares larger than that are
canonically called super-flares.

\begin{figure}
\includegraphics[height=0.35\textheight,angle=-90.0]{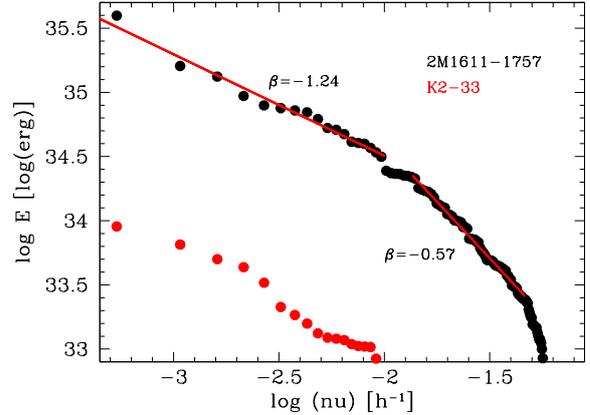}
\caption{Cumulative flare frequency distribution for 2M1611-1757
  (black points) and of K2-33 (red points). The solid lines represent
  our linear least-squares fit to the upper and lower part of the
  distribution.}
\label{Fig09}
\end{figure}

Because flares have a power-law distribution the cumulative frequency
distribution can be fitted using the relation:

\begin{equation}
{\rm log (\nu) = \gamma + \beta log(E)}
\label{eq01}
\end{equation}

where $\rm \nu$ is the cumulative frequency of flares with an energy greater 
than E and $\rm \beta$ is the power-law exponent 
\footnote{The power law distribution can also be described as $\rm
  dn/dE=k\,E^{\alpha}$. The relation between $\alpha$ and $\beta$ is:
  $\beta=1-\alpha$.}.  Figure~\ref{Fig09} shows the cumulative
frequency distribution for 2M1611-1757.  The red points in
Fig.~\ref{Fig09} is the cumulative frequency distribution of K2-33.
There is one flare $\rm E\geq 3\,10^{33} erg$ every 372
hours. Unfortunately, only 26 flares were detected in this star which
is not enough to determine the statistics of flares precisely.
However, there are enough flares to determine the ratio between the
flares in K2-33 and 2M1611-1757. Counting only flares with $\rm
E_{opt}\geq 10^{33}$\,erg, we find that flares in 2M1611-1757 emit 50
times the amount of energy as those in K2-33.

The smallest flare detected in 2M1611-1757 had $\rm
8\,10^{32}\,erg$. Since the completeness limit is three times higher
than the detection limit, we set a lower energy limit ($\rm E_{min}$)
down to which the flare-statistic is still complete, before we can
calculate the index.

Using $\rm E_{min}=2.6\,10^{33}\,erg$, we derive $\rm
\beta=-0.80\pm0.16$ for the whole distribution. A better way to
determine $\rm \beta$ is the maximum likelihood estimation
\citep{gizis2017}.  Using the same energy limit as above, we obtain
$\rm \beta=-0.65\pm0.15$.

As can be seen in Fig.~\ref{Fig09} 2M1611-1757 has a broken power-law
distribution. The energy range between log(E)=34.5 to 35.6 has a power
index $\rm \beta=-1.24$.  The energy range between log(E)=33.4 to 34.4
has $\rm \beta=-0.57$.  The significance of a broken power-law is
discussed in Section~\ref{sectVIk}.

\subsection{The frequency of flares for M-stars Upper Sco}
\label{sectVc}

\begin{figure}
\includegraphics[height=0.26\textheight,angle=0.0]{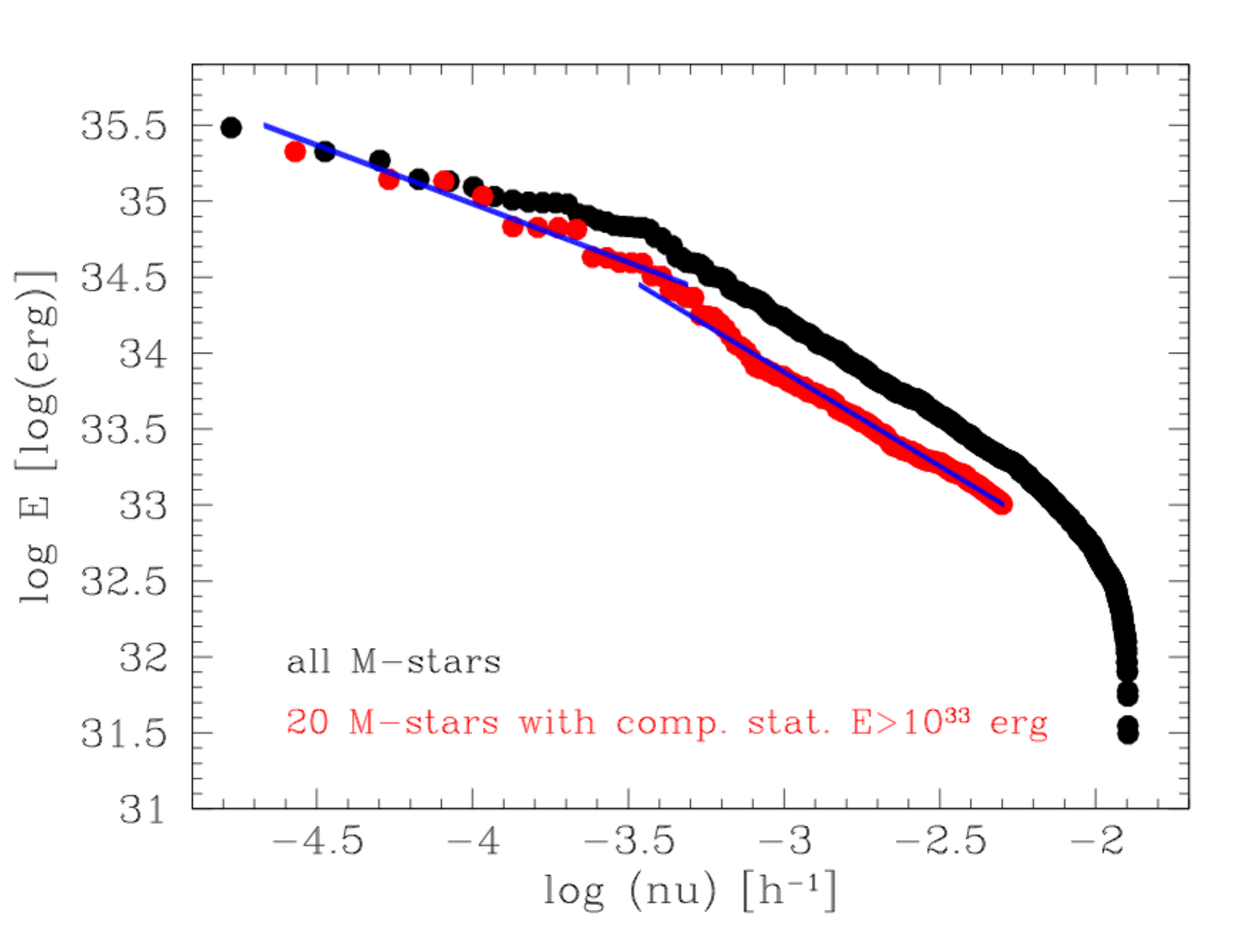}
\caption{Cumulative flare frequency distribution of all confirmed
M-stars in Upper Sco (black), and only for M-stars where the
completeness limit is better than $\rm 10^{33}\,erg$ (red points).}
\label{Fig12}
\end{figure}

Is 2M1611-1757 representative for an M-star at that age? To find out,
we determine the flare-statistics for all confirmed M-stars in Upper
Sco listed in Table~3.  The cumulative frequency distribution for all
confirmed M-stars, except 2M1611-1757, is shown in Fig.~\ref{Fig12}
(black points). We exclude 2M1611-1757 from this analysis, because we
would like to compare the cumulative frequency distribution of all
other M-stars with that of 2M1611-1757.

Using the maximum likelihood estimation, we obtain $\rm
\beta=-0.52\pm0.13$. The $\rm \beta$-values of 2M1611-1757 within the
errors are the same as that of all other M-stars in Upper Sco.

However, as pointed out by \cite{shibata2013} the detection threshold
for flares depends on the rotation period of the star.  The reason is
that smaller flares can more easily be detected in slowly rotating
stars than in rapidly rotating ones.  The detection threshold
furthermore depends on the signal-to noise of the light-curve, and the
brightness of the star. The last point is important because the
relative brightness increase of a flare is larger if the star is
intrinsically fainter. \citet{okamoto2021} introduced the detection
completeness filter $\rm DC_{filter}$ to correct for the missing
flares.  $\rm DC_{filter}$ is the ratio of stars where a flare of a
specific energy can certainly be detected to the total number of stars
in the sample.

\begin{figure}
\includegraphics[height=0.27\textheight,angle=0.0]{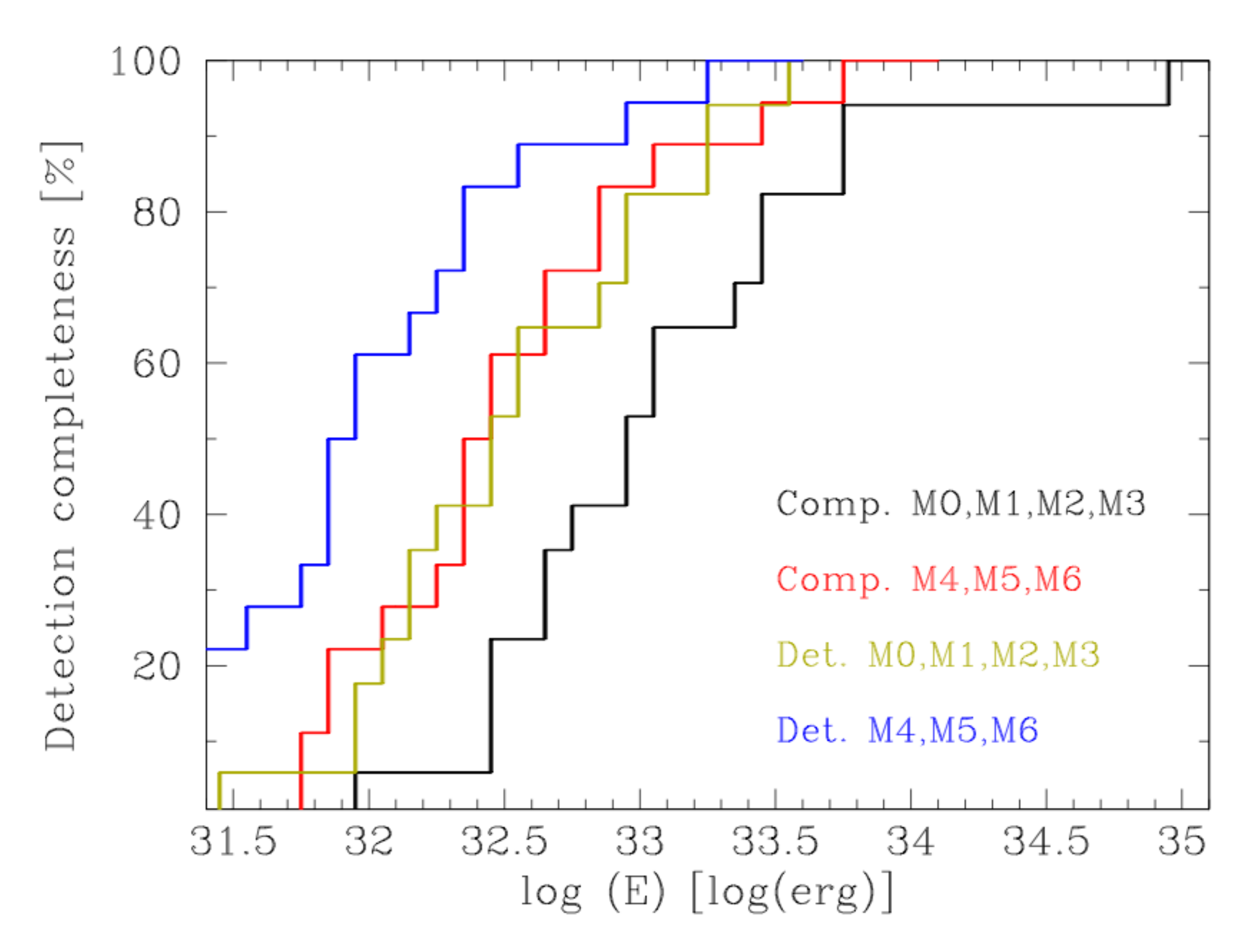}
\caption{Completeness and detection limits. Shown is the
fraction of stars where the detection and the completeness limit is
lower than the energy shown. "Det." is the detection limit
and "Comp" the completeness limit.}
\label{Fig13}
\end{figure}

We calculate for each star the detection limit for flares, and the
completeness limit. The detection limit is the energy of the smallest
flare observed in that star. The completeness limit is the energy down
to which all flares can be detected in that star. We find that the
main limiting factor for the flare detection in our sample is the
brightness of the star.  For this reason, we made the statistics for
the M0 to M3-stars, and for M4 to M6 stars separately.  We split the
sample in this way, so that both samples contain about the same number
of stars. Figure~\ref{Fig13} shows the detection and the completeness
limits for M-stars in Upper Sco.

Let us take energy of $\rm 10^{33}\,erg$ as an example.  Flares of
that energy can be detected in 82\% of the M0 to M3-stars.
However, all flares down to that energy can only be detected
in 52\% of the stars. For M0 to M3-stars, the detection and
completeness limits are 94\% and 82\% of the stars, respectively.

There are two ways to correct for the incompleteness. We can either
correct the flare-rates, or we can simply select only those stars
where the statistics down to $\rm 10^{33}\,erg$ are complete.  The red
points in Fig.~\ref{Fig12} show the cumulative frequency diagram for
the 20 M-stars where the completeness limit is better than $\rm
10^{33}$\,erg.  This means we can be certain that we have detected all
flares down to that energy in these stars.

\begin{figure}
\includegraphics[height=0.25\textheight,angle=0.0]{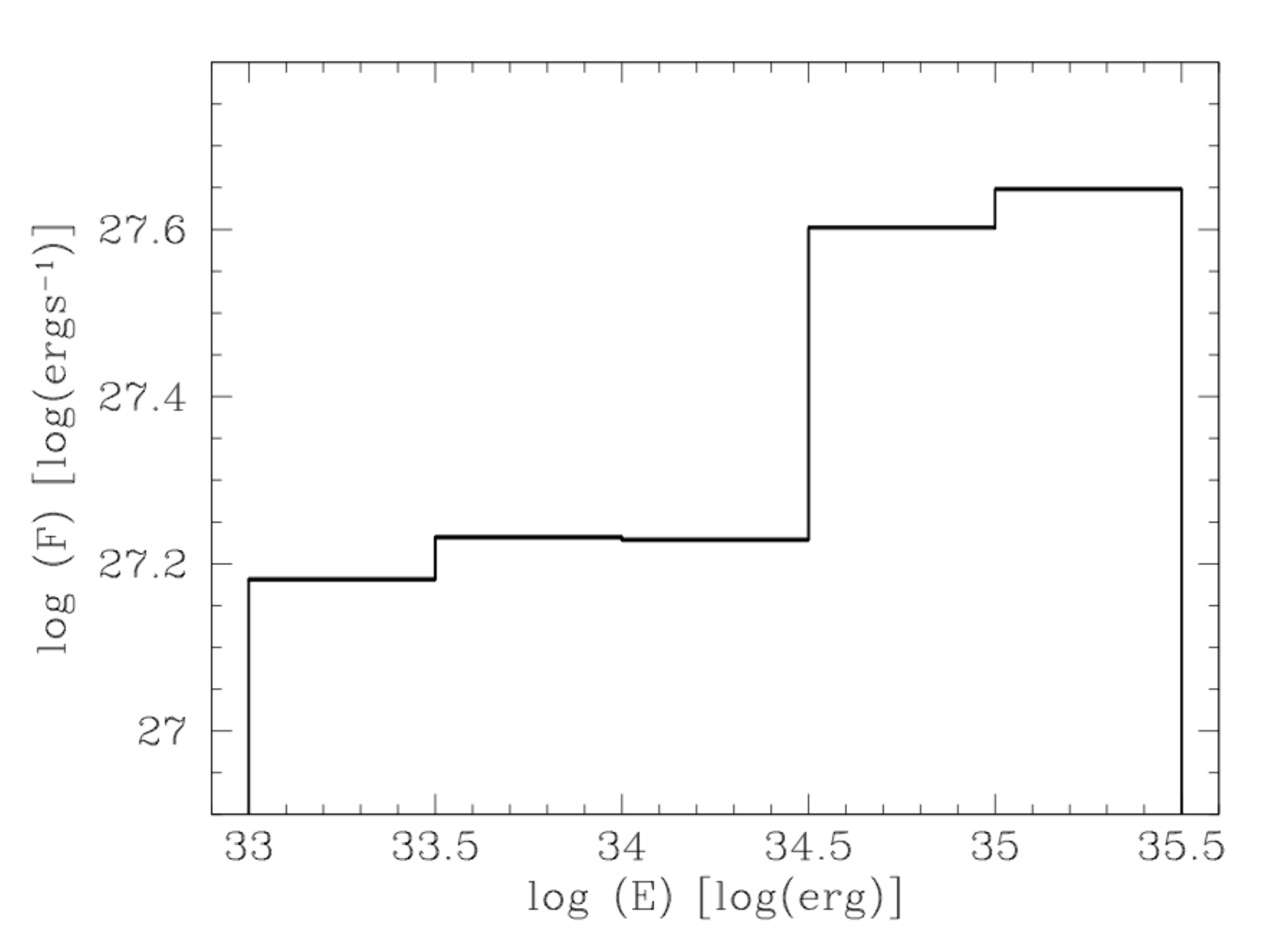}
\caption{Average amount of energy emitted by all flares in the energy 
intervals shown. For this analysis we used only the 20 M-stars in 
Upper Sco where all flares with $\rm E\ge10^{33}$ erg were detect.}
\label{Fig02}
\end{figure}

The flare-distribution turns out to be a broken power-law.  Flares
with $\rm E \geq 3\,10^{34}\,erg$ have $\rm \beta=-1.3\pm0.1$ and
flares smaller than that $\rm \beta=-0.8\pm0.1$.  Figure~\ref{Fig02}
shows the average amount of energy emitted per second by flares of
different energies. We used only the 20 M-stars where the
flare-statistics is complete down to $\rm 10^{33}\,erg$.  Per energy
interval more energy is emitted by larger flares than by small
ones. The dividing line for M-stars in upper Sco is $\rm E \geq
3\,10^{34}\,erg$.

\subsection{The decay time of the flares}
\label{sectVd}

\begin{figure}
\includegraphics[height=0.27\textheight,angle=0.0]{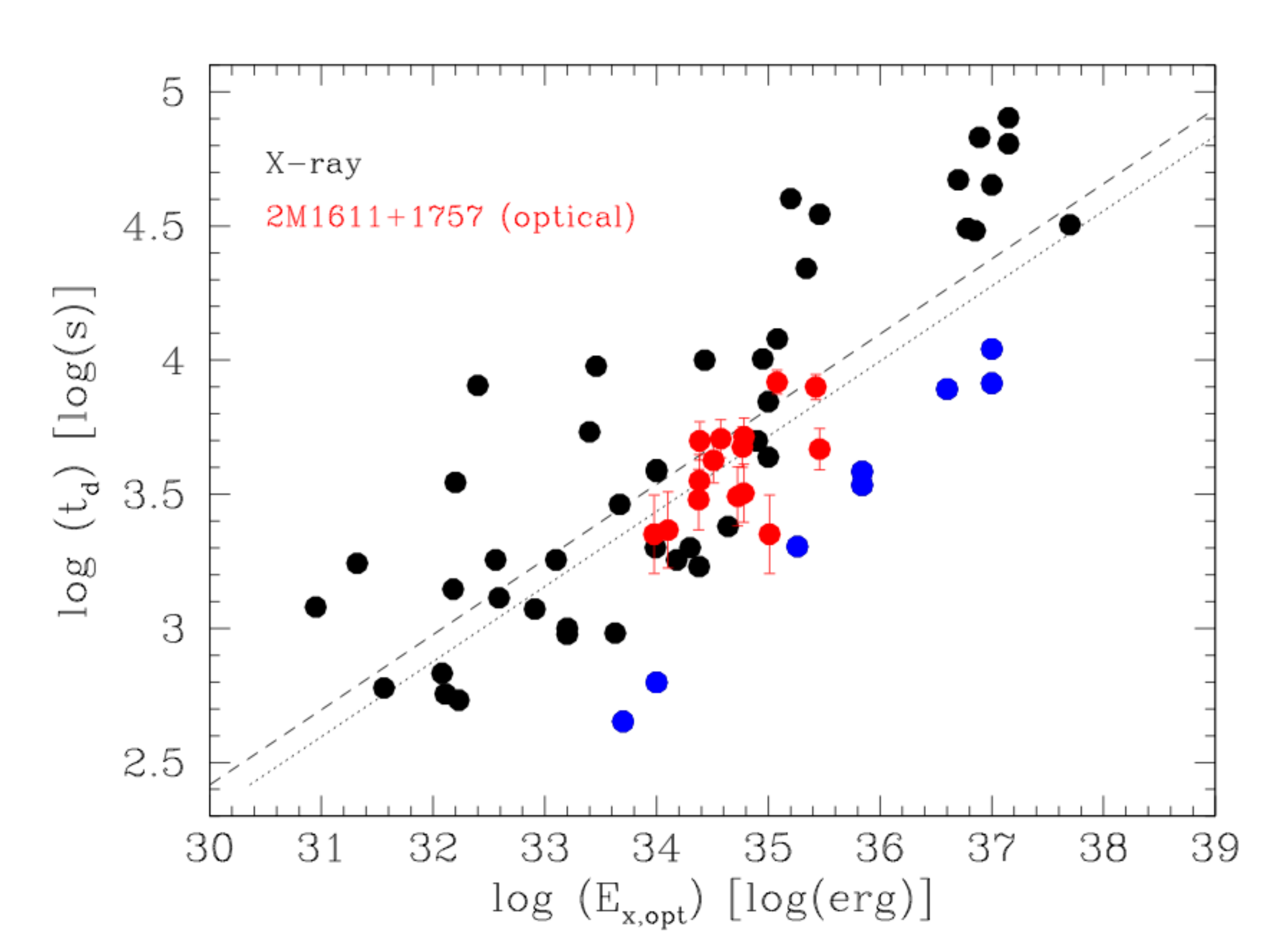}
\caption{Energy emitted by flares in the X-ray regime \citep{guedel04} 
  and in 2M1611-1757 in the optical regime versus the decay-time. 
  The upper dashed curve is for the X-ray data, the lower dotted, 
  the optical data.The blue { points are stars named in the text}. }
\label{Fig10}
\end{figure}

Since the energy of flares decreases exponentially, the decay-time
$\rm t_d$ of a flare is defined as the time at which the energy has
decreased by 1/e of its peak value. Figure \ref{Fig10} shows the
relation between the decay-time of the flares and the energy emitted.
The red points are for the optical continuum emission of
2M1611-1757. The black points are values obtained in the soft X-ray
regime for flares of active stars of different spectral types taken
from \citet{guedel04}.

We selected for this analysis only 15 flares with $\rm E_{opt}\geq
10^{34}\,erg $ that show a clean exponential decay. We did not include
multiple events overlapping each other. Although the black points is
the energy of flares emitted in soft X-rays and the red points the
energy emitted in the optical regime, the two fall almost on top of
each other. This result can be used to estimate the branching ratio of
flares between the optical and the soft X-rays in
Section~\ref{sectVIa}.

The relation between the decay time $\rm log(t_d)$ (log(s)) and the
energy of a flare in the soft X-ray regime $\rm log (E_x) $ (log
(erg)) is:

\begin{equation}
\rm log (t_d) = (-6.0\pm0.9) + (0.28\pm0.02) \cdot log (E_x) 
\label{eq02}
\end{equation}

The relation between the duration for flares and their energies for
super-flares on solar-like stars published by \citet{maehara2015} is
\footnote{The factor -10 is approximate. It is not given 
in the article, we determined it from the figure in that article.}: 

\begin{equation}
\rm log (t_d) = (\sim -10) + (0.39\pm0.03) \cdot log (E_{opt}) 
\label{eq03}
\end{equation}

Flare decay times for solar X-ray flares have been studied by
\citet{veronig2002} who found $\rm log (t_{d\,total}) \sim -0.96 \cdot
log (E_x)$, with $\rm t_{d\,total}$ the total duration of the flare
which includes the rise and decay time.  For solar hard/soft X-ray
flares, \citet{namekata2017} found $\rm log(t_d) \sim 0.2-0.33 \cdot
log(E_{opt})$. These factors equation~\ref{eq02} and \ref{eq03} are
surprisingly similar. Further discussions on that matter can be found
in Section~\ref{sectVIa}.

\section{Discussion}
\label{sectVI}

\subsection{Comparing $\rm \beta$ for young M-stars with with other stars}
\label{sectVIk}

We determined the power-law index of the flare-frequency distribution
for 2M1611-1757 in Section~\ref{sectVb}, and for all other young
M-stars in Upper Sco in Section~\ref{sectVc}.  M-stars in Upper Sco
have a broken power-law distribution with a critical energy of $\rm E
\geq 3\,10^{34}\,erg$. Larger flares have $\rm \beta=-1.3\pm0.1$ and
smaller ones $\rm \beta=-0.8\pm0.1$.

Broken power-laws have been reported for other stars as well.  For
example, a broken power-law was also observed in the giant star KIC
2852961 which has $\rm 1.7\pm0.3$ $\rm M_\odot$
\citep{kovari2020}. Flares with energies above $\rm E \geq 5\,10^{37}$
on this star have $\rm \beta=-1.84\pm0.06$ and flares below that
energy have $\rm \beta=-0.29\pm0.02$.

Why do stars have broken power-laws? A possible scenario was proposed
by \citet{mullan2018}.  Granules force the field lines at the
foot-points of coronal magnetic loops to undergo a random walk which
can twist the field lines so that a flare occurs. A critical point in
this scenario is the ratio between the diameter of the foot-points of
the loops to the magnetic scale height.  The frequency is different
for flares originating from loops with foot-points that are larger
than the scale height than for flares where the loops are smaller.  If
this hypothesis is correct, the flare frequency distribution should be
a broken power-law for all types stars.  This means if we observe a
star for a long time but our measurements are not particularly
sensitive, we will find $\rm \beta \leq -1$. The same result will be
obtained if the star is so inactive that we simply do not observe
really large flares.  If we observe a star for a short time but the
measurements are very sensitive, we will obtain $\rm \beta \geq
-1$. If we observe the same star over a long period, we may find that
the distribution is a broken power-law.

For example, $\rm \beta < -1$ were obtained by \citet{ilin2020} for
stars in open clusters, by \citet{shibayama2013} for super-flares of
G-dwarfs and by \citet{stelzer2007} for pre-main-sequence solar
analogies in the Taurus. White-light flares on the sun, and thus
presumably also on other solar-like stars, are large events which are
typically only observed in impulsive and energetic flares
\citep{watanabe2017}.  $\rm \beta > -1$ were, for example, obtained by
\citet{shibayama2013} for small solar flares, by \citet{jess2019} for
nano-flares on the sun, by \citet{audard2000} for flares on M-star,
and by \citet{maehara2015} for solar-like stars using the Kepler
short-cadence data.

All these values do not contradict each other if we assume that 
the underlying distribution is a broken power-law. In this context
it is interesting to note that a broken power-law can be seen in their
figure in the article published by \citet{stelzer2007}.

\subsection{The energy of the largest flare and the magnetic flux}
\label{sectVIj}

According to \citep{shibata2013}, the energy released in the largest
flares is related to the area of the star-spots. Because
\citep{shibata2013} studied solar-like stars, it would be interesting
to find out if this relation also holds for M-stars. The lower limit
of spot-filling factor of 2M1611-1757 is $\rm 7.1\pm0.9$ \%,
corresponding to $\rm 7.1\pm1.2\,10^{21}\,cm^{2}$ (see
Section\,\ref{sectVa}). Using the relation given \citep{shibata2013},
we derive an upper limit of the flare X-ray intensity of $\rm
2.5\pm0.7\,10^{36}\,erg$. The largest flare observed had only $\rm
4\,10^{35}\,erg$, and was thus one order of magnitude weaker than the
upper limit. The relation published by \citep{shibata2013} thus also
holds for M-stars.

Using the relationship between total unsigned magnetic flux ($\rm
\Phi$), and X-ray spectral radiance ($\rm L_x$) published by
\citet{pevtsov2003}, and $\rm L_x =10^{29}\, erg\,s^{-1}$ we estimate
$\rm \Phi=2.6\,10^{25}\,Mx$ for 2M1611-1757.  The magnetic flux
density thus estimated to be about $\rm B=3700\,G$.  The size of the
spots, the size of the flares and the magnetic flux of 2M1611-1757 are
thus similar to other active stars.  For example, the magnetic flux
density of the active M-star AD\,Leo is $\rm B=3300\,G$
\citep{cranmer2011}.

\subsection{Estimating the branching ratio between the optical 
and  the soft X-rays and the flare-decay time}
\label{sectVIa}

As pointed out by \citet{mullan2018} the energies released in the
optical continuum emission and the X-rays are related if the common
physical process contributes to both. This is the case if the energy
released in the coronal loops, which emit X-rays, is transported to
the photosphere from which optical emission emerges. If the optical
continuum emission is related to the X-ray emission, their decay times
should be related.

Observations of solar flares in fact show that the soft-X ray light
curve closely follows the $\rm H\alpha$-light curve
\citep{leitzinger2020}, and the white-light curve.  
If we assume that flares that have the same decay time are the same, we can
statistically derive the branching ratio between the optical regime
and the soft X-ray regime by shifting the optical points into position
of the X-ray points.

Using observations in the X-ray regime, and optical
photometry and spectroscopy of AD\,Leo,  \citet{namekata2020} 
determined a relation between the optical continuum and H$\alpha$ 
in the form $\rm I_{H\alpha} \sim I_{cont}^{0.51\pm0.05}$. 
One flare was observed with all instruments. 
This flare emitted $1.1\times 10^{30}$\,erg
in $\rm H\alpha$ and $3.4\times 10^{31}$\,erg in the 
(0.5 -10 keV) X-ray  regime. The light curves 
in the X-ray and  $\rm H\alpha$ have some similarity, but it appears 
that the decay time is longer in $\rm H\alpha$ than in X-rays.
In other events is also appears that the decay time is longer
in $\rm H\alpha$ than in optical continuum. Unfortunately, the optical 
continuum emission of this particularly flare was rather weak which makes
it difficult to compare the X-ray and optical continuum emission.

Figure~\ref{Fig10} shows the decay times for optical continuum
emission for M-stars in Upper Sco (red points), and the decay times
for active stars in the X-ray regime \citep{guedel04} (black and blue
points).  The relation of the decay time and the energy in the X-rays
and in the optical is given in Equation~\ref{eq02} and
Equation~\ref{eq03}, respectively.  Since the duration of flares is
related to the length scale (L) and the magnetic field strength (B) in
the form $\rm E \sim B^2\,L^3$ \citep{namekata2017, toriumi2017} it is
different for different types of stars.  This explains why the
equation~\ref{eq02} and \ref{eq03} are not identical.  Because the
black and blue points in Fig.~\ref{Fig10} are various types of stars,
the scatter has to be large. We marked as blue points the stars at the
lower edge of the distribution. These stars that are Castor, AD\,Dor
(K0V), EQ1839.6+8002 (M4V), AR\,Lac (K0IVe+G5IV) , II\,Peg (K2+IVe),
V773\,Tau (K3Ve) and Lk\,H$\rm \alpha$\,92 (G8e).

Nevertheless, if we assume that flares of the same decay time are
similar, we can estimate the branching ratio between the X-rays and
the optical continuum emission (Fig.~\ref{Fig10}). However, we should
keep in mind that the branching ratio depends on the energy of the
flare in the form $\rm F_{opt} \sim F_{x}^{0.59\pm0.04}$
\citep{namekata2017}.  That means the branching ratio depends on the
energy of the flare and the type of the star.  For flares on M-stars
in the energy-range from $10^{34}$ to $10^{36}$\,erg, we estimate a
branching-ratio of the order $\rm E_{opt}/E_{x}=1.5$ to 2.0.
\citet{osten2015} derived $\rm E_{Kepler}/E_{GOES}=3$ and $\rm
E_{Kepler}/E_{SRX}=0.5$.  The wavelength-ranges of these are: GOES:
1-8 \AA\, (1.55-12.4 keV), and SXR: 1.24-1240 \AA\, (10 eV-10
keV). The wavelength range of GOES thus is closer to normal soft-X
regime.  \citet{moore2014} found for X-Class Solar Flares that the
ratio of the total energy to the energy emitted in the 10-1900\,\AA \,
wavelength regime is $\rm E_{tot}/E_{FUV+EUV}=4.7\pm1.5$.

\subsection{What is the typical brightness increase of the
star due to flares?}
\label{sectVIb}

The total energy of all flares detected on 2M1611-1757 is $\rm
E_{opt}=2.2\,10^{36}$\,erg. Summing only the flares with $\rm
E_{opt}\geq 10^{34}$\,erg the total energy is $\rm
E_{opt}=1.9\,10^{36}$\,erg.  On average the flares with $\rm
E_{opt}\geq 10^{34}$\,erg enhance the optical emission by $\rm
L_{opt}\geq 3\,10^{29}$ $\rm erg\,s^{-1}$. Using the branching-ratio
from above, we estimate that the soft X-ray emission due to large
flares is of the order $\rm L_{x}\geq 2\,10^{29}$ $\rm erg\,s^{-1}$.
Since, the X-ray brightness of 2M1611-1757 is $\rm L_x=2,10^{30}$ $\rm
erg\,s^{-1}$.  However, we have to be careful with this conclusion,
because the scatter of the X-ray points in Fig. \ref{Fig10} is an
order of magnitude.  We need to observe flares in young M-stars
simultaneously in the optical continuum and the XUV.

\subsection{The duty cycle}
\label{sectVIe}

Defining the duration of flares is difficult, because of the
exponential decay. We thus define the duration of a flare as twice the
1/e time, or the time when the energy of the flare has decreased to
13.5\% of the peak energy.  We observed 40 flares with $\rm
E_{opt}\geq 10^{34}\,erg $ in 2M1611-1757.  If in this very active
star the duty cycle of such flares is only 4.6\%.

\subsection{The mass-loss rate of planets due to the flare activity}
\label{sectVIf}

\cite{morlock2020} used the results of this work in his model of the
erosion of planetary atmospheres. In their work, they derive a mass
loss of $\rm \dot{M}_\mathrm{eva} \simeq
3.22\times10^{-7}~\mathrm{M_\oplus}~\mathrm{yr}^{-1}$ for a
hypothetical 5 $\rm M_{Earth}$ planet orbiting 2M1611-1757 at 0.1
AU. This means that a H/He-envelope of a close-in mini-Neptune would
be removed. In that article they also derive a relation between the
EUV-flux from the star and the mass loss rate of the planet as $\rm
\Dot{M}_\mathrm{eva}\;\propto\;F_\mathrm{EUV}^{0.43\pm0.01}$.  This
relation allows us to calculate the mass-loss rate for planets
orbiting stars of different activity level.

As discussed in Section~\ref{sectVIb}, flares enhance the soft X-ray
emission of 2M1611-1757 by $\rm L_{x}\geq 2\,10^{29}$ $\rm
erg\,s^{-1}$. Since K2-33 is 50 times less active, we estimate $\rm
L_{x}\geq 4\,10^{27}$ $\rm erg\,s^{-1}$ for this star.  The erosion
rate due to flares alone would already be $\rm \dot{M}\sim
10^{-7}~\mathrm{M_\oplus}~\mathrm{yr}^{-1}$ if K2-33b had a mass of
five $\rm M_{\oplus}$.  This means if a H/He-envelope contains about
one percent of the mass of the planet, it would be substantially
eroded during the first few 100 Myrs due the XUV-radiation from flares
alone.  Flares thus play a role for the evolution of planetary
atmospheres, and should not be neglected.

\subsection{The erosion of planetary atmospheres is large 
due to the XUV-radiation, CMEs are perhaps less important}
\label{sectVIg}

In many studies it is assumed that the ratio of flares to
Coronal-Mass-Ejections (CMEs) is the same for active stars as for the
sun \citep{howard2018, yamashiki2019}. However, M-stars are
quite different from solar-like stars. It is thus not obvious that
we can assume the same flare-to-CMEs ratio as in solar-like stars.

CMEs are important, because the protons from CMEs significantly effect
the ozone layer of an Earth-like planet whereas the electromagnetic
radiation from flares effects the ozone layer to a much lesser extend.
\citet{tilley2019}, for example, estimates that proton events from
CMEs would deplete the ozone column of an Earth-like planet by 94\% in
only 10 years.  To reach this conclusion, they assumed a much lower
flare-rate than what we observed for 2M1611-1757. They assumed that
there is one flare of $\rm E_{U}\geq 10^{34} erg$ in the U-band every
11700 hours and a power-law index $\rm \beta=-1.01$. This corresponds
to a rate of one flare of $\rm E\geq 10^{34} erg$ every 8862 hours in
the Kepler band.

However, what do we actually know about the CME-activity of M-stars?
In our previous study of the very active M-star AD Leo, we obtained
2000 high resolution spectra and observed 22 flares
\citep{muheki2020a}. Line-asymmetries were often seen, but a
blue-shifted components exceeding the escape velocity of $\rm
590\pm11\,km\,s^{-1}$ were not observed. In another study we obtained
762 spectra of EV Lac. Again 27 flares were observed, also a filament
eruption but no CME \citep{muheki2020b}. Filament eruptions were also
observed by \citet{maehara2021} in YZ CMi. The velocity
of the blue-shifted component was between 80 and 100 $\rm km\,s^{-1}$.
The authors estimate the mass and kinetic energy of the upward-moving
material to be $\rm 10^{16}$ -- $10^{18}$\,g and $10^{29.5}$ --
$10^{31.5}$ erg, respectively.  The authors furthermore point out that
the kinetic energy of these events is two orders of magnitude lower
than that expected from the empirical relation for CMEs for our Sun.
\citet{fuhrmeister2018} obtained 473 high-resolution spectra with
CARMENES of 28 active M-dwarfs and detected 41 flares.
Line-asymmetries were again observed but the corresponding velocities
again did not reach escape velocities.  Doppler-shifted emission
features with shifts well below the the escape velocities were also
observed at FUV and X-ray wavelengths \citep{leitzinger2011,
  argiroffi2019}.  The same results were obtained in a dedicated
surveys for CMEs \citep{leitzinger2014, korhonen2017,leitzinger2020}.

The observations suggests that filament eruptions have been observed
but the velocities associated with them are lower than the escape
velocities from the stars, but this does not exclude the possibility
that there are CMEs.  The reason is that there are three stages of the
dynamic evolution of CMEs: a slow rise, a fast acceleration, and a
propagation phase \citep{gou2020}. At the end of the first phase solar
CMEs have reached a height of 0.5 $\rm R_\odot$, and at the end of the
second acceleration phase a height of 6 $\rm R_\odot$. Strictly
speaking, the events observed in H$\alpha$ are filament, or prominence
eruptions, not CMEs. While both phenomena are associated, the average
height of solar prominence eruption is only 0.36 $\rm R_\odot$
\citep{gopalswamy2003}.  This means that it is unlikely that we
measure the final speed of the ejected material if we obtain the
events in $\rm H\alpha$. Observations by \citet{gopalswamy2003} show
that the average speed measured in solar prominence eruption is only
56 $\rm km\,s^{-1}$.  Using the data presented in that article it
turns out that the median of the measured velocities of CMEs are an
order of magnitude higher than the velocities measured in the
prominence eruption associated with them. The relatively low speeds
observed in H$\alpha$ thus do not rule out CMEs.

However, \citet{odert2017} and \citet{drake2016} pointed out that the
mass-loss rates of active stars become unrealistically high if we
simply extrapolate the relation of the flare-energies to the mass-loss
rate for events that are orders a magnitude higher than solar
flares. As explained in these articles, it is plausible that a
relatively strong magnetic fields overlying the flaring region may
prevent matter from leaving the star. The flow patterns during flares
can be quite complex. For example, an up-flow of material at
temperature of $10^{4}$\,K that lifts up plasma with a temperature of
$10^{6-7}$\,K followed by downward moving condensation with a
temperature of $10^{4}$\,K has been observed on the sun
\citep{tei2018}. In the case of spatially resolved solar observations,
the full 3D-velocities of erupting filaments can be reconstructed and
these can be related to CMEs.  Solar observations show CMEs are in
general related to eruptive filaments that originate from flares that
emit more than $\rm 3\times10^{25}\,ergs^{-1}$ in the 1-8\AA \, regime
\citep{morimoto2003}.  However, if only one spectral line is observed,
it is difficult to say if this event will cause a CME, or not.
Multi-wavelength observations and observations with the next
generation of radio telescopes may solve the CME-mystery
\citep{osten2018}.

In summary, observations in H$\alpha$ show only
filament eruptions not CMEs. It is possible that these are associated with CMEs
but that remains to be shown. However, there are a number of 
reasonable arguments why the CME-rate of M-stars is perhaps 
not that high. The hypothesis of a relatively low CME-rate is consistent 
with the results obtained by \citet{wood2014}.
Using  an absorption feature in $\rm Ly\alpha$ line as a tracer for 
mass-loss,   \citet{wood2014} find that the loss increases up to a certain
activity level, but above that the mass loss is relatively small. Since
most of the mass loss is due to CMEs \citep{linsky2015b}, this
also means that the CME-activity of very active stars is relatively low.

2M1611-1757 thus may also be in the weak-wind regime. Using the
results obtained by \citet{wood2014}, the 
mass loss is expected to be of the order of $\rm \dot{M}\sim
2\,10^{-13}\,M_\odot\,year^{-1}$.  Putting everything together, young 
M-stars have CMEs but the CME-activity is not as spectacular as the 
flare activity. 

\subsection{The role of flares for the habitability of planets}
\label{sectVIi}

In a detailed study of the evolution of the activity of stars
\citet{johnstone2020}, showed that although the rates of flares at all
energies are higher for G dwarfs, but the amount of XUV-radiation that
a planet in the habitable zone receives from flares are likely to be
higher for M dwarfs. More in detail, for $\rm \beta=-0.6$, the authors
find that the amount is roughly the same for all stars in the mass
range between 0.2 and 1.4 $\rm M_\odot$ but for $\rm \beta=-1.4$
planets in the habitable zone of M-stars receive two three orders of
magnitude more energy from flares. Planets in the habitable zone of
M-stars receive two, or three orders of magnitude more energy from
energetic flares than planets in the habitable zone of G-stars.

In this respect it is interesting to note that the isotope ratios $\rm
^{36}Ar/^{38}Ar$, $\rm ^{20}Ne/^{22}Ne$, and $\rm ^{36}Ar/^{22}Ne$ on
Earth and Venus can only be explained if it is assumed that the young
Sun was only weakly active in the first 100 Myrs
\citep{lammer2020}. Analyzing the sodium and potassium of the lunar
regolith, \citet{saxena2019} come to the same conclusions: The young
Sun was particularly inactive even compared to solar-like stars, and
even more so compared to M-stars.

Flares not only enhance the amount of XUV-radiation, they also have a
different X-ray spectrum. This is demonstrated by the observations of
the young, active K0-star AB\,Dor.  Observations of this star show
that the flux of 20 MK component increases significantly during
flares, whereas the 7 and 3 MK components remain basically constant
\citep{besselaar2003}.  This higher temperature of flares is
important, because the heating efficiency of planetary atmospheres
depends on the XUV-spectrum of the star and thus the temperature of
the emitting region \citep{shematovich2014}. The UV radiation controls
the photochemical reactions in planetary atmospheres for important
molecules such as $\rm H_2O$, $\rm CO_2$, and $\rm CH_4$
\citep{linsky2015b}.

Since flares significantly increase the amount of this type of
radiation, they are highly important for the photochemistry of
planetary atmospheres \citep {lammer2018}.  Biological studies
demonstrate the vulnerability of microorganisms if they are exposed to
the UV-C radiation (1000-2800 \AA) from super-flares
\citep{abrevaya2020}.  Thus, flares can make planets uninhabitable
even if they are formally in the habitable zone.  Perhaps it is not a
chance coincidence that the Earth is orbiting a G-star rather than an
M-star.

\section{Conclusions}
\label{sectVII} 

Using spectroscopic and photometric observations, we have studied the
flare activity of 2MASS\,J1611-1757.  This star is particularly
interesting because it is perhaps the first M-star in which solar-like
oscillations have been discovered \citep{muellner2020}. We find that
2MASS\,J1611-1757 is a member of the Upper Sco OB association of young
stars. The analysis of the K2 light curves shows that it has an
enormous flare activity. There is one flare with $\rm E\geq
10^{35}$\,erg every 620 hours, and one with $\rm E\geq 10^{34} erg$
every 52 hours.  This star is an ideal target for studying the
activity of young M-stars.

The flare-energy distribution of young M-stars in Upper Sco is a
broken power-law. Large flares have $\rm \beta=-1.3\pm0.1$ and smaller
ones $\rm \beta=-0.8\pm0.1$. The critical point between the two
regimes is at $\rm E \geq 3\,10^{34}$\,erg. The broken power law
distribution naturally explains why some previous studies obtained
values smaller than -1 and others values larger than that.  Surveys in
which only large flares are detected obtained values $\rm
\beta\leq-1$.  Sensitive observations obtained over a short time
interval give $\rm \beta\geq-1$. Sensitive, long-duration surveys will
find a broken power-law.  The distribution of the decay times is
similar to that of other active stars, and also not that different
from that of solar-like stars.

We monitored the star spectroscopically for 10.5 hours but did not
detect a flare, a CME, nor the signature of an eruptive filament.  We
conclude that flares are important for the evolution of planetary
atmospheres and should not be neglected.

\section*{Acknowledgements}

Data Availability Statement: The data underlying this article will be
shared on reasonable request to the corresponding author. The FLAMES
spectra are public and can be obtained from the ESO data-archive.  The
FLAMES-UVES observations were obtained in ESO programe
097.C-0040(A). This work was generously supported by the Deutsche
Forschungsgemeinschaft (DFG) in the framework of the priority programe
"Exploring the Diversity of Extrasolar Planets" (SPP 1992) in program
GU 464/22, in the DFG-program HA3279/11-1, and by the Th\"uringer
Ministerium f\"ur Wirtschaft, Wissenschaft und Digitale Gesellschaft.
VK thanks PRL for his PhD research fellowship.  The research work at
the Physical Research Laboratory is funded by the Department of Space,
Government of India.  We are very thankful to the ESO-staff for
carrying out the observations in service mode, and for providing the
community with all the necessary tools for reducing and analyzing the
data.  This research has made use of the SIMBAD database, operated at
CDS, Strasbourg, France.

This work has made use of data from the European Space Agency (ESA)
mission {\it Gaia} (\url{https://www.cosmos.esa.int/gaia}), processed
by the {\it Gaia} Data Processing and Analysis Consortium (DPAC,
\url{https://www.cosmos.esa.int/web/gaia/dpac/consortium}). Funding
for the DPAC has been provided by national institutions, in particular
the institutions participating in the {\it Gaia} Multilateral
Agreement.  The Gaia mission website is
\url{https://www.cosmos.esa.int/gaia}
https://www.cosmos.esa.int/gaia. The Gaia archive website is
\url{https://archives.esac.esa.int/gaia.}  This publication makes use
of data products from the Wide-field Infrared Survey Explorer, which
is a joint project of the University of California, Los Angeles, and
the Jet Propulsion Laboratory/California Institute of Technology,
funded by the National Aeronautics and Space Administration.

We would like to thank the referee for the suggestions that
  helped to improve the article significantly.

\section{Apendix: Upper Sco members}

Remarks: \\
$^1$ Rotation period [d], $^2$ Kepler magnitude, $^3$ Gaia magnitude, 
$^4$ Gaia early data release 3 (\url{https://gea.esac.esa.int/archive/}\citep{gaia16, gaia18, gaia20}, 
$^5$ Spectrum has a S/N of 2-8 which is too low to measure the EW. $^6$ visual binary in Gaia DR3, 
$^7$ unrelated star within 3 arcsec in Gaia DR3,
$^8$ K2-33, data from \citep{mann2016}.

\begin{sidewaystable*}[ht]
\centering
  \begin{tabular}{l l r c c c c c c c c c}  
\hline
EPIC & ST & No.    & $\rm P_{rot}^1$ &  RA    &  DEC   & $\rm Kp^2$    & $\rm G^3$     & $\rm plx^4$   & $\rm p.m.^4$     & $\rm p.m.^4$    & EW  \\
     &    & fla-   & [d]       & h:m:s  & d:m:s & [mag] & [mag] & [mas] & RA       & DEC     & LiI\,6707 \\
     &    & res    &           &        &       &       &       &       & [mas/a]  & [mas/a] & [\AA ] \\
\hline
204813678 & M4 & 12  &  2.31 & 16:07:16.072 & -20:44:43.80 & 12.97 & 14.74 & $7.029\pm0.037$  & $-11.522\pm0.042$ & $-23.247\pm0.028$ & $0.593\pm0.008$ \\  
204829482 & M5 & 12  &  1.95 & 16:08:00.517 & -20:40:28.93 & 14.87 & 15.95 & $6.464\pm0.062$  & $ -9.392\pm0.081$ & $-22.356\pm0.050$ & $0.634\pm0.060$ \\  
204832041 & M6 & 5   &  0.58 & 16:07:58.508 & -20:39:48.59 & 14.93 & 17.56 & $6.442\pm0.123$  & $ -9.391\pm0.169$ & $-23.157\pm0.105$ & $\rm low S/N^5$ \\  
204838119 & M3 & 14  &  3.98 & 16:08:15.353 & -20:38:11.19 & 16.21 & 14.89 & $6.579\pm0.034$  & $ -9.821\pm0.046$ & $-22.439\pm0.027$ & $0.512\pm0.003$ \\  
204845955 & M4 & 6   &  1.85 & 16:07:44.490 & -20:36:02.94 & 13.54 & 13.32 & $6.863\pm0.067$  & $ -7.668\pm0.083$ & $-25.142\pm0.051$ & $0.567\pm0.006$ \\  
204854345 & K5 & 5   &  5.32 & 16:08:56.731 & -20:33:45.87 & 11.66 & 11.61 & $7.050\pm0.017$  & $ -9.035\pm0.022$ & $-25.209\pm0.015$ & $0.487\pm0.004$ \\  
204876697 & K8 & 28  &  9.28 & 16:08:01.414 & -20:27:41.66 & 13.08 & 12.96 & $7.051\pm0.019$  & $-10.671\pm0.022$ & $-22.646\pm0.016$ & $0.536\pm0.008$ \\  
204895521 & M2 & 33  &  4.89 & 16:06:47.509 & -20:22:32.18 & 14.01 & 13.84 & $7.228\pm0.023$  & $-10.579\pm0.027$ & $-23.029\pm0.018$ & $\rm low S/N^5$ \\  
204901273 & M3 & 62  &  4.97 & 16:07:19.722 & -20:20:55.55 & 14.14 & 15.00 & $6.399\pm0.034$  & $ -8.520\pm0.040$ & $-22.122\pm0.027$ & $0.392\pm0.025$ \\  
204906020$^6$ & M5 & 1 & - & 16:07:02.118 & -20:19:38.77 & 16.99 & 16.55 & $6.818\pm0.099$ & $-10.621\pm0.119$ & $-21.361\pm0.081$ & $0.294\pm0.078$ \\  
204909952 & M5 & 0   & -     & 16:07:27.545 & -20:18:34.44 & 16.96 & 16.41 & $7.061\pm0.081$  & $-10.836\pm0.099$ & $-23.252\pm0.070$ & $\rm low S/N^5$ \\  
204919503 & M4 & 19  &  3.11 & 16:07:04.742 & -20:15:55.75 & 16.13 & 15.88 & $6.453\pm0.051$  & $ -9.823\pm0.064$ & $-21.774\pm0.041$ & $0.363\pm0.024$ \\  
204948308 & M5 & 22  &  0.90 & 16:02:10.961 & -20:07:49.58 & 14.91 & 15.83 & $6.690\pm0.049$  & $ -9.455\pm0.056$ & $-21.839\pm0.032$ & $0.599\pm0.027$ \\  
204966558 & M5 & 17  &  2.17 & 16:02:26.138 & -20:02:40.71 & 14.41 & 15.35 & $6.634\pm0.043$  & $ -9.610\pm0.052$ & $-23.123\pm0.031$ & $0.606\pm0.021$ \\  
204986988 & M3 & 12  &  1.68 & 16:02:22.481 & -19:56:53.96 & 14.94 & 14.88 & $6.603\pm0.033$  & $-10.571\pm0.041$ & $-21.947\pm0.024$ & $0.558\pm0.007$ \\  
204993463 & M5 & 1   &  3.42 & 16:03:29.416 & -19:55:03.64 & 14.54 & 15.31 & $6.390\pm0.043$  & $ -9.362\pm0.051$ & $-20.904\pm0.035$ & $0.477\pm0.006$ \\  
205014939 & M5 & 7   &  0.55 & 16:04:07.756 & -19:48:57.77 & 15.72 & 16.41 & $10.555\pm0.069$ & $-43.179\pm0.074$ & $-21.014\pm0.049$ & $\rm low S/N^5$ \\  
205026365 & M3 & 25  &  2.03 & 16:10:10.416 & -19:45:39.81 & 13.91 & 14.53 & $7.178\pm0.030$  & $-11.322\pm0.038$ & $-22.825\pm0.025$ & $0.493/-0.003$  \\  
205034152 & M3 & 20  &  3.62 & 16:09:13.392 & -19:43:28.15 & 14.12 & 14.73 & $7.226\pm0.038$  & $-8.915\pm0.039$  & $-24.118\pm0.030$ & $0.521\pm0.012$ \\  
205038557 & M2 & 69  &  3.83 & 16:03:57.931 & -19:42:10.91 & 14.48 & 14.30 & $6.486\pm0.028$  & $-11.270\pm0.032$ & $-22.096\pm0.020$ & $0.410\pm0.005$ \\  
205040772$^7$ & M3 & 5 &  1.12 & 16:10:21.526 & -19:41:31.86 & 14.22 & 14.28 & $6.227\pm0.70$ & $-7.683\pm0.304$ & $-23.567\pm0.211$ & $0.553\pm0.029$ \\  
205046529$^6$ & M4 & 8 & 12.83 & 16:10:26.386 & -19:39:51.06 & 13.73 & 15.52 & $7.091\pm0.096$ & $-8.765\pm0.110$ & $-24.316\pm0.072$ & $0.474\pm0.030$ \\  
205054397 & M3 & 21  & 1.810 & 16:10:52.406 & -19:37:34.45 & 14.24 & 15.12 & $7.165\pm0.036$ &  $-8.292\pm0.048$ & $-24.096\pm0.032$ & $0.374\pm0.009$ \\  
205056060 & M4 & 27  & 11.33 & 16:09:15.808 & -19:37:06.30 & 15.99 & 15.73 & $6.432\pm0.047$ &  $-7.805\pm0.056$ & $-21.317\pm0.041$ & $0.593\pm0.025$ \\  
205060410 & M2 & 39  & 17.96 & 16:08:23.870 & -19:35:51.91 & 13.06 & 13.14 &     -           &       -           &       -           & $0.553\pm0.002$ \\  
205066998$^6$ & M5 & -   & - & 16:09:04.054 & -19:34:00.05 & 16.00 & 15.93 & $6.316\pm0.239$ & $-9.975\pm0.288$  & $-19.022\pm0.201$ & $0.507\pm0.057$ \\  
205080616 & K9 & 0   & 10.7  & 16:08:23.244 & -19:30:00.96 & 13.22 & 13.13 & $7.256\pm0.023$ & $-12.437\pm0.024$ & $-22.477\pm0.021$ & $0.531\pm0.003$ \\  
205089832 & M2 & 86  & 8.19  & 16:07:07.676 & -19:27:16.23 & 13.79 & 13.73 & $6.456\pm0.085$ & $-10.088\pm0.124$ & $-20.683\pm0.082$ & $0.495\pm0.053$ \\  
205109605$^6$ & M0 & 2 & -   & 16:08:28.662 & -19:21:24.46 & 15.25 & 15.53 & $6.089\pm0.043$ & $-14.202\pm0.049$ & $-28.847\pm0.033$ & $\rm low S/N^5$ \\  
205117205$^8$ & $\rm M3.3\pm0.5$ & 26 & $\rm 6.29\pm0.17$ & 16:10:14.738 &  -19:19:09.40 & 14.36 & 14.13 & $7.193\pm0.023$ & $-9.592\pm0.028$ & $-23.964\pm0.021$ & $0.45\pm0.15$ \\
205124839$^7$ & M2 & 35 & 4.70 & 16:10:39.569 & -19:16:52.45 & 14.71 & 14.54 & $6.401\pm0.026$ & $-9.036\pm0.032$ & $-21.112\pm0.026$ & $0.481\pm0.006$ \\  
205133963 & M3 & 5   & 1.58  & 16:10:24.740 & -19:14:07.35 & 12.96 & 14.63 & $6.444\pm0.031$ & $-8.955\pm0.042$ & $-21.213\pm0.030$ & $0.525\pm0.035$ \\  
205137430 & K7 & 44  & 12.31 & 16:10:31.951 & -19:13:06.07 & 12.78 & 12.69 & $7.293\pm0.041$ & $-8.632\pm0.055$ & $-23.634\pm0.038$ & $0.542\pm0.005$ \\  
205142641 & M2 & 98  & 6.30  & 16:08:23.572 & -19:11:31.60 & 14.29 & 14.06 & $7.267\pm0.026$ & $-8.424\pm0.029$ & $-24.674\pm0.020$ & $0.521\pm0.005$ \\  
205145188 & M4 & 2   & -     & 16:10:28.196 & -19:10:44.49 & 17.00 & 16.61 & $6.592\pm0.067$ & $-9.471\pm0.087$ & $-21.632\pm0.064$ & $0.396\pm0.029$ \\  
205151387 & M4 & 2   & -     & 16:09:00.763 & -19:08:52.70 & 13.29 & 12.89 & $7.278\pm0.024$ & $-9.374\pm0.028$ & $-25.121\pm0.019$ & $0.470\pm0.007$ \\  
205152244 & M5 & 18  & 1.81  & 16:09:00.036 & -19:08:37.62 & 13.42 & 15.42 & $7.303\pm0.052$ & $-9.439\pm0.060$ & $-24.891\pm0.040$ & $\rm low S/N^5$ \\  
205164832 & K3 & 4   & 3.06  & 16:08:10.828 & -19:04:47.94 & 11.12 & 11.27 & $7.195\pm0.023$ & $-9.104\pm0.022$ & $-24.910\pm0.015$ & $0.450\pm0.002$ \\  
205164892 & M3 & 25  & 6.65  & 16:10:28.571 & -19:04:46.90 & 12.90 & 12.81 & -               & -                & -                 & $0.560\pm0.002$ \\  
205177770 & M4 & 1   & 1.14  & 16:08:43.098 & -19:00:51.88 & 16.14 & 14.62 & -               & -                & -                 & $0.542\pm0.020$ \\  
205355375 & M6 & 34  & 1.16  & 16:11:18.211 & -18:03:58.55 & 15.13 & 16.70 & $7.174\pm0.088$ & $-8.495\pm0.115$ & $-24.108\pm0.084$ & $\rm low S/N^5$ \\  
205375290 & M1 & 138 & 6.03  & 16:11:15.342 & -17:57:21.44 & 13.23 & 12.92 & $7.365\pm0.017$ & $-8.954\pm0.021$ & $-24.626\pm0.016$ & $0.588\pm0.002$ \\  
   &  &  &  &  &  &  &  &   &  &   & \\
   &  &  &  &  &  &  &  &   &  &   & \\
   &  &  &  &  &  &  &  &   &  &   & \\
   &  &  &  &  &  &  &  &   &  &   & \\
   &  &  &  &  &  &  &  &   &  &   & \\
   &  &  &  &  &  &  &  &   &  &   & \\
   &  &  &  &  &  &  &  &   &  &   & \\
   &  &  &  &  &  &  &  &   &  &   & \\
   &  &  &  &  &  &  &  &   &  &   & \\
   &  &  &  &  &  &  &  &   &  &   & \\
   &  &  &  &  &  &  &  &   &  &   & \\
   &  &  &  &  &  &  &  &   &  &   & \\
   &  &  &  &  &  &  &  &   &  &   & \\
   &  &  &  &  &  &  &  &   &  &   & \\
   &  &  &  &  &  &  &  &   &  &   & \\
   &  &  &  &  &  &  &  &   &  &   & \\
   &  &  &  &  &  &  &  &   &  &   & \\
   &  &  &  &  &  &  &  &   &  &   & \\
   &  &  &  &  &  &  &  &   &  &   & \\
   &  &  &  &  &  &  &  &   &  &   & \\
   &  &  &  &  &  &  &  &   &  &   & \\
   &  &  &  &  &  &  &  &   &  &   & \\
   &  &  &  &  &  &  &  &   &  &   & \\
   &  &  &  &  &  &  &  &   &  &   & \\
   &  &  &  &  &  &  &  &   &  &   & \\
   &  &  &  &  &  &  &  &   &  &   & \\
   &  &  &  &  &  &  &  &   &  &   & \\
   &  &  &  &  &  &  &  &   &  &   & \\
   &  &  &  &  &  &  &  &   &  &   & \\
   &  &  &  &  &  &  &  &   &  &   & \\
   &  &  &  &  &  &  &  &   &  &   & \\
   &  &  &  &  &  &  &  &   &  &   & \\
   &  &  &  &  &  &  &  &   &  &   & \\
   &  &  &  &  &  &  &  &   &  &   & \\
   &  &  &  &  &  &  &  &   &  &   & \\
   &  &  &  &  &  &  &  &   &  &   & \\
   &  &  &  &  &  &  &  &   &  &   & \\
   &  &  &  &  &  &  &  &   &  &   & \\
   &  &  &  &  &  &  &  &   &  &   & \\
   &  &  &  &  &  &  &  &   &  &   & \\
   &  &  &  &  &  &  &  &   &  &   & \\
   &  &  &  &  &  &  &  &   &  &   & \\
   &  &  &  &  &  &  &  &   &  &   & \\
   &  &  &  &  &  &  &  &   &  &   & \\
   &  &  &  &  &  &  &  &   &  &   & \\
   &  &  &  &  &  &  &  &   &  &   & \\
   &  &  &  &  &  &  &  &   &  &   & \\
   &  &  &  &  &  &  &  &   &  &   & \\
\hline
\end{tabular}
\label{tab02}
\end{sidewaystable*}





\begin{thebibliography}{99}




\bibitem[\protect\citeauthoryear{Abrevaya et al.}{2020}]{abrevaya2020}
  Abrevaya X.~C., Leitzinger M., Oppezzo O.~J., Odert P., Patel M.~R.,
  Luna G.~J.~M., Forte Giacobone A.~F., et al., 2020, MNRAS, 494,
  L69. doi:10.1093/mnrasl/slaa037

\bibitem[\protect\citeauthoryear{Argiroffi et
    al.}{2019}]{argiroffi2019} Argiroffi C., Reale F., Drake J.~J.,
  Ciaravella A., Testa P., Bonito R., Miceli M., et al., 2019, NatAs,
  3, 742. doi:10.1038/s41550-019-0781-4

\bibitem[\protect\citeauthoryear{Audard et al.}{2000}]{audard2000}
  Audard M., G{\"u}del M., Drake J.~J., Kashyap V.~L., 2000, ApJ, 541,
  396. doi:10.1086/309426

\bibitem[\protect\citeauthoryear{Baraffe et al.}{2015}]{baraffe2015}
  Baraffe I., Homeier D., Allard F., Chabrier G., 2015, A\&A, 577,
  A42. doi:10.1051/0004-6361/201425481

\bibitem[\protect\citeauthoryear{Bohn et al.}{2020}]{bohn2020} Bohn
  A.~J., Kenworthy M.~A., Ginski C., Rieder S., Mamajek E.~E., Meshkat
  T., Pecaut M.~J., et al., 2020, ApJL, 898,
  L16. doi:10.3847/2041-8213/aba27e

\bibitem[\protect\citeauthoryear{Cloutier \&
    Menou}{2020}]{cloutier2020} Cloutier R., Menou K., 2020, AJ, 159,
  211. doi:10.3847/1538-3881/ab8237

\bibitem[\protect\citeauthoryear{Cranmer \& Saar}{2011}]{cranmer2011}
  Cranmer S.~R., Saar S.~H., 2011, ApJ, 741,
    54. doi:10.1088/0004-637X/741/1/54

\bibitem[\protect\citeauthoryear{Crespo-Chac{\'o}n et
    al.}{2006}]{crespo2006} Crespo-Chac{\'o}n I., Montes D.,
  Garc{\'\i}a-Alvarez D., Fern{\'a}ndez-Figueroa M.~J.,
  L{\'o}pez-Santiago J., Foing B.~H., 2006, A\&A, 452,
  987. doi:10.1051/0004-6361:20053615

\bibitem[\protect\citeauthoryear{Cutri et al.}{2013}]{cutri2013} Cutri
  R.~M., Wright E.~L., Conrow T., Fowler J.~W., Eisenhardt P.~R.~M.,
  Grillmair C., Kirkpatrick J.~D., et al., 2013, wise.rept

\bibitem[\protect\citeauthoryear{David et al.}{2019}]{david2019} David
  T.~J., Hillenbrand L.~A., Gillen E., Cody A.~M., Howell S.~B.,
  Isaacson H.~T., Livingston J.~H., 2019, ApJ, 872,
  161. doi:10.3847/1538-4357/aafe09

\bibitem[\protect\citeauthoryear{Drake et al.}{2016}]{drake2016} 
 Drake J.~J., Cohen O., Garraffo C., Kashyap V., 2016, 
IAUS, 320, 196. doi:10.1017/S1743921316000260

\bibitem[\protect\citeauthoryear{Fang, Herczeg, \&
    Rizzuto}{2017}]{fang2017} Fang Q., Herczeg G.~J., Rizzuto A.,
  2017, ApJ, 842, 123. doi:10.3847/1538-4357/aa74ca

\bibitem[\protect\citeauthoryear{Fridlund et al.}{2020}]{fridlund2020}
  Fridlund M., Livingston J., Gandolfi D., Persson C.~M., Lam
  K.~W.~F., Stassun K.~G., Hellier C., et al., 2020, MNRAS, 498,
  4503. doi:10.1093/mnras/staa2502
  
\bibitem[\protect\citeauthoryear{Fuhrmeister et
    al.}{2018}]{fuhrmeister2018} Fuhrmeister B., Czesla S., Schmitt
  J.~H.~M.~M., Jeffers S.~V., Caballero J.~A., Zechmeister M., Reiners
  A., et al., 2018, A\&A, 615, A14. doi:10.1051/0004-6361/201732204

\bibitem[\protect\citeauthoryear{Gaia Collaboration}{2016}]{gaia16}
  Gaia Collaboration, Prusti T., de Bruijne J.~H.~J., Brown A.~G.~A.,
  Vallenari A., Babusiaux C., Bailer-Jones C.~A.~L., et al., 2016,
  A\&A, 595, A1. doi:10.1051/0004-6361/201629272
 
\bibitem[\protect\citeauthoryear{Gaia Collaboration}{2018}]{gaia18}
  Gaia Collaboration, Brown A.~G.~A., Vallenari A., Prusti T., de
  Bruijne J.~H.~J., Babusiaux C., Bailer-Jones C.~A.~L., et al., 2018,
  A\&A, 616, A1
  
 \bibitem[\protect\citeauthoryear{Gaia Collaboration}{2020}]{gaia20}
  Gaia Collaboration, Gaia early data release 3,
  \url{https://archives.esac.esa.int/gaia.}

\bibitem[\protect\citeauthoryear{Giardino et al.}{2008}]{giardino2008}
  Giardino G., Pillitteri I., Favata F., Micela G., 2008, A\&A, 490,
  113. doi:10.1051/0004-6361:200810042

\bibitem[\protect\citeauthoryear{Garufi et al.}{2020}]{garufi20}
  Garufi A., Avenhaus H., P{\'e}rez S., Quanz S.~P., van Holstein
  R.~G., Bertrang G.~H.-M., Casassus S., et al., 2020, A\&A, 633, A82

\bibitem[\protect\citeauthoryear{Ginzburg, Schlichting, \&
    Sari}{2018}]{ginzburg2018} Ginzburg S., Schlichting H.~E., Sari
  R., 2018, MNRAS, 476, 759. doi:10.1093/mnras/sty290
  
\bibitem[\protect\citeauthoryear{Gizis et al.}{2017}]{gizis2017} Gizis
  J.~E., Paudel R.~R., Mullan D., Schmidt S.~J., Burgasser A.~J.,
  Williams P.~K.~G., 2017, ApJ, 845, 33. doi:10.3847/1538-4357/aa7da0

\bibitem[\protect\citeauthoryear{G{\"u}del}{2004}]{guedel04} G{\"u}del
  M., 2004, A\&ARv, 12, 71. doi:10.1007/s00159-004-0023-2

\bibitem[\protect\citeauthoryear{Guenther et al.}{2017}]{guenther17}
  Guenther E.~W., Barrag{\'a}n O., Dai F., Gandolfi D., Hirano T.,
  Fridlund M., Fossati L., et al., 2017, A\&A, 608,
  A93. doi:10.1051/0004-6361/201730885

\bibitem[\protect\citeauthoryear{Gupta \&
    Schlichting}{2019}]{gupta2019} Gupta A., Schlichting H.~E., 2019,
  MNRAS, 487, 24. doi:10.1093/mnras/stz1230

\bibitem[\protect\citeauthoryear{Gupta \&
    Schlichting}{2020}]{gupta2020} Gupta A., Schlichting H.~E., 2020,
  MNRAS, 493, 792. doi:10.1093/mnras/staa315

\bibitem[\protect\citeauthoryear{Gopalswamy et al.}{2003}]{gopalswamy2003} 
Gopalswamy N., Shimojo M., Lu W., Yashiro S., Shibasaki K., 
Howard R.~A., 2003, ApJ, 586, 562. doi:10.1086/367614

\bibitem[\protect\citeauthoryear{Gou et al.}{2020}]{gou2020} 
Gou T., Veronig A.~M., Liu R., Zhuang B., Dumbovi{\'c} M., Podladchikova T., 
Reid H.~A.~S., et al., 2020, ApJL, 897, L36. doi:10.3847/2041-8213/ab9ec5

\bibitem[\protect\citeauthoryear{Houdebine, Foing, \&
    Rodono}{1990}]{houdebine1990} Houdebine E.~R., Foing B.~H., Rodono
  M., 1990, A\&A, 238, 249
 
\bibitem[\protect\citeauthoryear{Howard et al.}{2018}]{howard2018}
  Howard W.~S., Tilley M.~A., Corbett H., Youngblood A., Loyd
  R.~O.~P., Ratzloff J.~K., Law N.~M., et al., 2018, ApJL, 860,
  L30. doi:10.3847/2041-8213/aacaf3

\bibitem[\protect\citeauthoryear{Hunt-Walker et
    al.}{2012}]{huntwalker2012} Hunt-Walker N.~M., Hilton E.~J.,
  Kowalski A.~F., Hawley S.~L., Matthews J.~M., 2012, PASP, 124,
  545. doi:10.1086/666495

\bibitem[\protect\citeauthoryear{Ilin et al.}{2020}]{ilin2020} Ilin
  E., Schmidt S.~J., Poppenh{\"a}ger K., Davenport J.~R.~A.,
  Kristiansen M.~H., Omohundro M., 2020, arXiv, arXiv:2010.05576

\bibitem[\protect\citeauthoryear{Jackson \&
    Jeffries}{2013}]{jackson2013} Jackson R.~J., Jeffries R.~D., 2013,
  MNRAS, 431, 1883. doi:10.1093/mnras/stt304

\bibitem[\protect\citeauthoryear{Jess et al.}{2019}]{jess2019} Jess
  D.~B., Dillon C.~J., Kirk M.~S., Reale F., Mathioudakis M., Grant
  S.~D.~T., Christian D.~J., et al., 2019, ApJ, 871,
  133. doi:10.3847/1538-4357/aaf8ae

\bibitem[\protect\citeauthoryear{Jin et al.}{2014}]{jin2014} Jin S.,
  Mordasini C., Parmentier V., van Boekel R., Henning T., Ji J., 2014,
  ApJ, 795, 65. doi:10.1088/0004-637X/795/1/65

\bibitem[\protect\citeauthoryear{Johnstone et
    al.}{2015}]{johnstone2015} Johnstone C.~P., G{\"u}del M., Brott
  I., L{\"u}ftinger T., 2015, A\&A, 577,
  A28. doi:10.1051/0004-6361/201425301

\bibitem[\protect\citeauthoryear{Johnstone, Bartel, \&
    G{\"u}del}{2020}]{johnstone2020} Johnstone C.~P., Bartel M.,
  G{\"u}del M., 2020, arXiv, arXiv:2009.07695

\bibitem[\protect\citeauthoryear{Korhonen et al.}{2017}]{korhonen2017}
  Korhonen H., Vida K., Leitzinger M., Odert P., Kov{\'a}cs O.~E.,
  2017, IAUS, 328, 198. doi:10.1017/S1743921317003969

\bibitem[\protect\citeauthoryear{K{\H{o}}v{\'a}ri et
    al.}{2020}]{kovari2020} K{\H{o}}v{\'a}ri Z., Ol{\'a}h K.,
  G{\"u}nther M.~N., Vida K., Kriskovics L., Seli B., Bakos G. {\'A}.,
  et al., 2020, A\&A, 641, A83. doi:10.1051/0004-6361/202038397

\bibitem[\protect\citeauthoryear{Kubyshkina et
    al.}{2018}]{kubyshkina2018} Kubyshkina D., Lendl M., Fossati L.,
  Cubillos P.~E., Lammer H., Erkaev N.~V., Johnstone C.~P., 2018,
  A\&A, 612, A25. doi:10.1051/0004-6361/201731816

\bibitem[\protect\citeauthoryear{Lacy, Moffett, \&
    Evans}{1976}]{lacy1976} Lacy C.~H., Moffett T.~J., Evans D.~S.,
  1976, ApJS, 30, 85. doi:10.1086/190358

\bibitem[\protect\citeauthoryear{Lammer et al.}{2014}]{lammer2014}
  Lammer H., St{\"o}kl A., Erkaev N.~V., Dorfi E.~A., Odert P.,
  G{\"u}del M., Kulikov Y.~N., et al., 2014, MNRAS, 439,
  3225. doi:10.1093/mnras/stu085

\bibitem[\protect\citeauthoryear{Lammer et al.}{2018}]{lammer2018}
  Lammer H., Zerkle A.~L., Gebauer S., Tosi N., Noack L., Scherf M.,
  Pilat-Lohinger E., et al., 2018, A\&ARv, 26,
  2. doi:10.1007/s00159-018-0108-y

\bibitem[\protect\citeauthoryear{Lammer et al.}{2020}]{lammer2020}
  Lammer H., Leitzinger M., Scherf M., Odert P., Burger C., Kubyshkina
  D., Johnstone C., et al., 2020, Icar, 339,
  113551. doi:10.1016/j.icarus.2019.113551

\bibitem[\protect\citeauthoryear{Leitzinger et
    al.}{2011}]{leitzinger2011} Leitzinger M., Odert P., Ribas I.,
  Hanslmeier A., Lammer H., Khodachenko M.~L., Zaqarashvili T.~V., et
  al., 2011, A\&A, 536, A62. doi:10.1051/0004-6361/201015985

\bibitem[\protect\citeauthoryear{Leitzinger et
    al.}{2014}]{leitzinger2014} Leitzinger M., Odert P., Greimel R.,
  Korhonen H., Guenther E.~W., Hanslmeier A., Lammer H., et al., 2014,
  MNRAS, 443, 898. doi:10.1093/mnras/stu1161

\bibitem[\protect\citeauthoryear{Leitzinger et
    al.}{2020}]{leitzinger2020} Leitzinger M., Odert P., Greimel R.,
  Vida K., Kriskovics L., Guenther E.~W., Korhonen H., et al., 2020,
  MNRAS, 493, 4570. doi:10.1093/mnras/staa504

\bibitem[\protect\citeauthoryear{Linsky \&
    G{\"u}del}{2015}]{linsky2015a} Linsky J.~L., G{\"u}del M., 2015,
  ASSL, 3. doi:10.1007/978-3-319-09749-7\,1

\bibitem[\protect\citeauthoryear{Lopez \& Fortney}{2014}]{lopez2014}
  Lopez E.~D., Fortney J.~J., 2014, ApJ, 792,
  1. doi:10.1088/0004-637X/792/1/1

\bibitem[\protect\citeauthoryear{Mann et al.}{2016}]{mann2016} Mann
  A.~W., Newton E.~R., Rizzuto A.~C., Irwin J., Feiden G.~A., Gaidos
  E., Mace G.~N., et al., 2016, AJ, 152,
  61. doi:10.3847/0004-6256/152/3/61

\bibitem[\protect\citeauthoryear{Maehara et al.}{2015}]{maehara2015}
 Maehara H., Shibayama T., Notsu Y., Notsu S., Honda S., Nogami
    D., Shibata K., 2015, EP\&S, 67,
    59. doi:10.1186/s40623-015-0217-z
  

\bibitem[\protect\citeauthoryear{Maehara et al.}{2021}]{maehara2021} 
Maehara H., Notsu Y., Namekata K., Honda S., Kowalski A.~F., Katoh N., 
Ohshima T., et al., 2021, PASJ, 73, 44. doi:10.1093/pasj/psaa098

\bibitem[\protect\citeauthoryear{Masuda}{2014}]{masuda14} Masuda K.,
  2014, ApJ, 783, 53. doi:10.1088/0004-637X/783/1/53

\bibitem[\protect\citeauthoryear{Moore, Chamberlin, \&
    Hock}{2014}]{moore2014} Moore C.~S., Chamberlin P.~C., Hock R.,
  2014, ApJ, 787, 32. doi:10.1088/0004-637X/787/1/32

\bibitem[\protect\citeauthoryear{Morlock et al.}{2020}]{morlock2020}
  Morlock, M., Kuiper, R., Guenther, E.W., W\"ockel, D., 20202, in
  preparation
  
 \bibitem[\protect\citeauthoryear{Morimoto \& Kurokawa}{2003}]{morimoto2003} 
 Morimoto T., Kurokawa H., 2003, PASJ, 55, 1141. doi:10.1093/pasj/55.6.1141
  
\bibitem[\protect\citeauthoryear{Muheki et al.}{2020a}]{muheki2020a}
  Muheki P., Guenther E.~W., Mutabazi T., Jurua E., 2020, A\&A, 637,
  A13. doi:10.1051/0004-6361/201936904

\bibitem[\protect\citeauthoryear{Muheki et al.}{2020b}]{muheki2020b}
  Muheki P., Guenther E.~W., Mutabazi T., Jurua E., 2020,
  MNRAS.tmp. doi:10.1093/mnras/staa3152

\bibitem[\protect\citeauthoryear{Mullan \& Paudel}{2018}]{mullan2018}
 Mullan D.~J., Paudel R.~R., 2018, ApJ, 854,
    14. doi:10.3847/1538-4357/aaa960

\bibitem[\protect\citeauthoryear{M\"ullner et
    al.}{2018}]{muellner2020} M\"ullner, M., Zwintz, K., Corsaro, E.,
  et al. 2020, submitted

\bibitem[\protect\citeauthoryear{Namekata et al.}{2017}]{namekata2017}
   Namekata K., Sakaue T., Watanabe K., Asai A., Maehara H., Notsu
    Y., Notsu S., et al., 2017, ApJ, 851,
    91. doi:10.3847/1538-4357/aa9b34
    
\bibitem[\protect\citeauthoryear{Namekata et al.}{2020}]{namekata2020} 
Namekata K., Maehara H., Sasaki R., Kawai H., Notsu Y., Kowalski
A.~F., Allred J.~C., et al., 2020, PASJ, 72,
68. doi:10.1093/pasj/psaa051


\bibitem[\protect\citeauthoryear{Notsu et al.}{2019}]{notsu2019} Notsu
  Y., Maehara H., Honda S., Hawley S.~L., Davenport J.~R.~A., Namekata
  K., Notsu S., et al., 2019, ApJ, 876,
  58. doi:10.3847/1538-4357/ab14e6

\bibitem[\protect\citeauthoryear{Odert et al.}{2017}]{odert2017} Odert
  P., Leitzinger M., Hanslmeier A., Lammer H., 2017, MNRAS, 472,
  876. doi:10.1093/mnras/stx1969

\bibitem[\protect\citeauthoryear{Odert et al.}{2020}]{odert2020} Odert
  P., Leitzinger M., Guenther E.~W., Heinzel P., 2020, MNRAS, 494,
  3766. doi:10.1093/mnras/staa1021

\bibitem[\protect\citeauthoryear{Okamoto et al.}{2021}]{okamoto2021}
  Okamoto S., Notsu Y., Maehara H., Namekata K., Honda S., Ikuta
    K., Nogami D., et al., 2021, ApJ, 906,
    72. doi:10.3847/1538-4357/abc8f5
 
\bibitem[\protect\citeauthoryear{Osten \& Wolk}{2015}]{osten2015}
  Osten R.~A., Wolk S.~J., 2015, ApJ, 809,
  79. doi:10.1088/0004-637X/809/1/79

\bibitem[\protect\citeauthoryear{Osten, Crosley, \& Hallinan}{2018}]{osten2018} 
Osten R.~A., Crosley M.~K., Hallinan G., 2018, ASPC, 517, 229

\bibitem[\protect\citeauthoryear{Owen \& Wu}{2013}]{owen2013} Owen
  J.~E., Wu Y., 2013, ApJ, 775, 105. doi:10.1088/0004-637X/775/2/105

\bibitem[\protect\citeauthoryear{Preibisch et al.}{2001}]{preibisch01}
  Preibisch, T., Guenther, E., Zinnecker, H., 2001, AJ, 121, 1040

\bibitem[\protect\citeauthoryear{Pettersen, Coleman, \&
    Evans}{1984}]{pettersen1984} Pettersen B.~R., Coleman L.~A., Evans
  D.~S., 1984, ApJS, 54, 375. doi:10.1086/190934

\bibitem[\protect\citeauthoryear{Pevtsov et al.}{2003}]{pevtsov2003}
  Pevtsov A.~A., Fisher G.~H., Acton L.~W., Longcope D.~W.,
  Johns-Krull C.~M., Kankelborg C.~C., Metcalf T.~R., 2003, ApJ, 598,
  1387. doi:10.1086/378944

\bibitem[\protect\citeauthoryear{Quirrenbach et al.}{2020}]{quirrenbach2020} 
Quirrenbach A., CARMENES Consortium, Amado P.~J., Ribas I., Reiners A., 
Caballero J.~A., Aceituno J., et al., 2020, SPIE, 11447, 114473C. doi:10.1117/12.2561380

\bibitem[\protect\citeauthoryear{Rajpurohit et
    al.}{2020}]{rajpurohit2020} Rajpurohit A.~S., Kumar V., Srivastava
  M.~K., Allard F., Homeier D., Dixit V., Patel A., 2020, MNRAS, 492,
  5844. doi:10.1093/mnras/staa163

\bibitem[\protect\citeauthoryear{Reid, Hawley, \&
    Gizis}{1995}]{reid1995} Reid I.~N., Hawley S.~L., Gizis J.~E.,
  1995, AJ, 110, 1838. doi:10.1086/117655
  
\bibitem[\protect\citeauthoryear{Saxena et al.}{2019}]{saxena2019}
  Saxena P., Killen R.~M., Airapetian V., Petro N.~E., Curran N.~M.,
  Mandell A.~M., 2019, ApJL, 876, L16. doi:10.3847/2041-8213/ab18fb

\bibitem[\protect\citeauthoryear{Shematovich, Ionov, \&
    Lammer}{2014}]{shematovich2014} Shematovich V.~I., Ionov D.~E.,
  Lammer H., 2014, A\&A, 571, A94. doi:10.1051/0004-6361/201423573

\bibitem[\protect\citeauthoryear{Shibata et al.}{2013}]{shibata2013} 
Shibata K., Isobe H., Hillier A., Choudhuri A.~R., Maehara H., Ishii T.~T.,
 Shibayama T., et al., 2013, PASJ, 65, 49. doi:10.1093/pasj/65.3.49

\bibitem[\protect\citeauthoryear{Shibayama et
    al.}{2013}]{shibayama2013} Shibayama T., Maehara H., Notsu S.,
  Notsu Y., Nagao T., Honda S., Ishii T.~T., et al., 2013, ApJS, 209,
  5. doi:10.1088/0067-0049/209/1/5

\bibitem[\protect\citeauthoryear{Shkolnik, Liu, \&
    Reid}{2009}]{shkolnik2009} Shkolnik E., Liu M.~C., Reid I.~N.,
  2009, ApJ, 699, 649. doi:10.1088/0004-637X/699/1/649

\bibitem[\protect\citeauthoryear{Skrutskie et al.}{2006}]{2MASS}
  Skrutskie, M.~F., Cutri, R.~M., Stiening, R., et al.\ 2006, \aj,
  131, 1163

\bibitem[\protect\citeauthoryear{Srivastava et
    al.}{2018}]{srivastava2018} Srivastava M.~K., Jangra M., Dixit V.,
  Munjal B.~S., Arora H., Mavani T., 2018, SPIE, 10702,
  107024I. doi:10.1117/12.2309306
  
 \bibitem[\protect\citeauthoryear{Srivastava et al.}{2021}]{srivastava2021} 
Srivastava M.~K., Kumar V., Dixit V., Patel A., Jangra M., Rajpurohit A.~S., 
 Mathur S.~N., 2021, arXiv, arXiv:2104.00314

\bibitem[\protect\citeauthoryear{Stelzer et al.}{2007}]{stelzer2007}
  Stelzer B., Flaccomio E., Briggs K., Micela G., Scelsi L., Audard
  M., Pillitteri I., et al., 2007, A\&A, 468,
  463. doi:10.1051/0004-6361:20066043

\bibitem[\protect\citeauthoryear{Tei et al.}{2018}]{tei2018} 
Tei A., Sakaue T., Okamoto T.~J., Kawate T., Heinzel P., UeNo S., 
Asai A., et al., 2018, PASJ, 70, 100. doi:10.1093/pasj/psy047

\bibitem[\protect\citeauthoryear{Telleschi et
    al.}{2005}]{telleschi2005} Telleschi A., G{\"u}del M., Briggs K.,
  Audard M., Ness J.-U., Skinner S.~L., 2005, ApJ, 622,
  653. doi:10.1086/428109 Telleschi, A., G\"udel, M., Briggs, K., et
  al. 2005, ApJ, 622, 653

\bibitem[\protect\citeauthoryear{Tilley et al.}{2019}]{tilley2019}
  Tilley M.~A., Segura A., Meadows V., Hawley S., Davenport J., 2019,
  AsBio, 19, 64. doi:10.1089/ast.2017.1794

\bibitem[\protect\citeauthoryear{Toriumi et al.}{2017}]{toriumi2017}
  Toriumi S., Schrijver C.~J., Harra L.~K., Hudson H., Nagashima
    K., 2017, ApJ, 834, 56.  doi:10.3847/1538-4357/834/1/56

\bibitem[\protect\citeauthoryear{Tsurutani et
    al.}{2003}]{tsurutani2003} Tsurutani B.~T., Gonzalez W.~D.,
  Lakhina G.~S., Alex S., 2003, JGRA, 108,
  1268. doi:10.1029/2002JA009504

\bibitem[\protect\citeauthoryear{van den Besselaar et
    al.}{2003}]{besselaar2003} van den Besselaar E.~J.~M., Raassen
  A.~J.~J., Mewe R., van der Meer R.~L.~J., G{\"u}del M., Audard M.,
  2003, A\&A, 411, 587. doi:10.1051/0004-6361:20031398

\bibitem[\protect\citeauthoryear{Voges et al.}{2000}]{ROSAT2000} Voges
  W., Aschenbach B., Boller T., Brauninger H., Briel U., Burkert W.,
  Dennerl K., et al., 2000, yCat, IX/29

\bibitem[\protect\citeauthoryear{Veronig et al.}{2002}]{veronig2002} 
Veronig A., Temmer M., Hanslmeier A., Otruba W., Messerotti M., 2002, 
A\&A, 382, 1070. doi:10.1051/0004-6361:20011694

\bibitem[\protect\citeauthoryear{Watanabe, Kitagawa, \& Masuda}{2017}]{watanabe2017}
Watanabe K., Kitagawa J., Masuda S., 2017, ApJ, 850, 204. 
doi:10.3847/1538-4357/aa9659

\bibitem[\protect\citeauthoryear{Wenger et al.}{2000}]{wenger00}
  Wenger, M., Ochsenbein, F., Egret, D., et al.\ 2000, \aaps, 143, 9

\bibitem[\protect\citeauthoryear{Wood et al.}{2014}]{wood2014} Wood
  B.~E., M{\"u}ller H.-R., Redfield S., Edelman E., 2014, ApJL, 781,
  L33. doi:10.1088/2041-8205/781/2/L33

\bibitem[\protect\citeauthoryear{Wood, Linsky, \&
    G{\"u}del}{2015}]{linsky2015b} Wood B.~E., Linsky J.~L., G{\"u}del
  M., 2015, ASSL, 19. doi:10.1007/978-3-319-09749-7\,2

\bibitem[\protect\citeauthoryear{Wood et al.}{2018}]{wood2018} Wood
  B.~E., Laming J.~M., Warren H.~P., Poppenh\"ager K., 2018, ApJ, 862,
  66. doi:10.3847/1538-4357/aaccf6

\bibitem[\protect\citeauthoryear{Yamashiki et
    al.}{2019}]{yamashiki2019} Yamashiki Y.~A., Maehara H., Airapetian
  V., Notsu Y., Sato T., Notsu S., Kuroki R., et al., 2019, ApJ, 881,
  114. doi:10.3847/1538-4357/ab2a71
  

\end{thebibliography}







\bsp	
\label{lastpage}
\end{document}